# Biodegradable drug eluting coating of cardiovascular stents dewets and can cause thrombosis

Alexey Kondyurin, Irina Kondyurina, Marcela Bilek
Applied and Plasma Physics, School of Physics (A28), The University of Sydney, New South Wales 2006, Australia, kond@mailcity.com

## Introduction

Biodegradable polymers are used as a matrix in drug-eluting coatings on current generation cardiovascular stents [1]. These polymers degrade in living organisms by a hydrolysis reaction leading to the release of drugs embedded in them. These drugs are effective in reducing inflammatory responses and thrombosis. Due to these advantages, the drug eluting stents found a wide application in vascular surgery. However, some studies have found that degradation of the matrix polymer also releases reaction products that cannot be metabolized safely and that are not retained in tissues. The question still remains whether drug- eluting stents show significant improvement compared to uncoated stents and is discussed [2, 3]. Because of these contradicting results, there are still doubts in the use of drug-eluting stents.
The disadvantages associated with the biodegradation process in polymer coating on metal surface may be a result of energy minimization between the polymer and metal surfaces. The surface energy of polymer coating is significantly less than the surface energy of a metal surface of the stent and these two substances will decrease the interface area following to the minimization of free energy. If the polymer coating is movable, a dewetting effect of polymer coating on stents surface is expected [4].

One of the successful candidates for biodegradable coatings is Poly(lactic-co-glycolic acid) (PLGA) which is used as a biodegradable polymer in medical applications [5-19]. PLGA degrades in living organisms through a hydrolysis reaction in which degradation products, such as lactic and glycolic acids, are released. The released products can be metabolized by the organism and are not toxic or collected in tissues. Due to its safe biodegradability, PLGA is used as a theraputic agent in organism as a substrate for drug delivery systems and for tissue engineering implants. PLGA is easy to deal with and a number of microspheres, microparticles, nanoparticles and coatings of different size and shape can be produced. PLGA coatings on stents degrade with time in organisms and can slowly release a drugs. Drugs encapsulated in PLGA can gradually and locally release during the biodegradation.

The kinetics of drug release from polymer coating is described generally by the so called “frontal process” of water diffusion into the polymer. According to this model, the release process is determined by the motion of the water front, which penetrates into the polymer film with time of exposition in blood stream. The rate of the “frontal process” depends on the water diffusion coefficient into the polymer film, the degradation rate of polymer in the water media and the solubility of the drug in water [20-22].

Effecting the “frontal process,” the drug release kinetics from the polymer coating are non-uniform. Initially, the drug release is high, resulting in large amounts of drug releases into the blood stream. Thereby, the polymer coating must be designed such that the amount of drug released during the initial period is below the toxic level. With time, the rate of drug release decreases and eventually the amount of released falls below the level of drug requisite for therapeutic effects, which limits the lifetime of the device.

In previous paper we showed, that ion beam treatment cans make the kinetics of drug release more uniform on polyurethane, which can increase its lifetime [23]. In this study, we investigate whether similar advantages can be found with PLGA coatings.

Ion beam implantation and its variant plasma immersion ion implantation (PIII) are based on collision and energy transfers from high energy ions penetrating into the polymer target. The depth of ion penetration depends on the kind of ion, its mass and the target density and atomic composition. Usually, the energy of the ion is much higher than the energy required to break chemical bonds in the polymer. Therefore, the penetrating ions break a number of bonds in macromolecules, sputter separate atoms and electrons from macromolecules and excite the vibrational states of the macromolecules. A new chemical structure is then formed after displaced atoms and electrons come to rest and the dissipation of thermal energy has occurred. A detail overview of the relevant physical and chemical processes can be found in [24-26].

Ion beam implantation and PIII were used for the modification of polyethylene [27-35], polystyrene [36-42], polytetrafluorethylene [43-48], polyethyleneterephtalat [49-52], polymethylmethacrylate [53-56], polyurethane [57-60] and other polymers. Structural changes, typical of ion irradiated polymers, such as carbonisation, depolymerisation and oxidation were observed, leading to increases in hardness, electrical conductivity and wettability. After ion implantation with sufficient fluence all polymers form a highly carbonized surface layer independent on the initial polymer structure.

This report presents a study of the PLGA structure after PIII and the degradation kinetics of PLGA coating in water after modification.

## Experiment

PLGA powder was received from Boston Scientific Inc. The powder, as received, was dissolved in acetone (Sigma Aldrich, Australia). The PLGA coating was spun from solution (model EC101 spin coater from Headway Research Inc, USA) onto (100) silicon wafers and steel polished plates. The nominal thickness of PLGA coating was 100 nm from 1 g/l concentration solutions, and 1 μm from 10 g/l concentration solutions. The coating was spun at 1500 rpm to achieve a homogenous thickness profile over the substrate surface of 1x1 $cm^2$ and 2x2 $cm^2$ size.

The PLGA coating spun at room temperature (23°C) is non-uniform due phase separation in the final stage of acetone evaporation. Some microstructures in spun PLGA coating are visible in optical microscope (Figure 1). The effect of phase separation was decreased with heating of the silicon wafer and solution up to 35-40°C. After spincoating, the PLGA samples were dried overnight. Thick PLGA films of 5x5 $cm^2$ were made on glass substrates (1-0.5 mm thickness of the film) and on polyester fabric (15-20 μm thickness of the film) from solution by casting. Thick films were dried in air at room temperature overnight and then placed in a vacuum chamber (1 torr) at room temperature for 4 h. The absence of acetone in the PLGA films was determined by FTIR spectra.

Vascular stents from Boston Scientific Inc. coated with PLGA were also modified and analyzed.

Plasma immersion ion implantation was conducted in an inductively coupled radio-frequency (13.56 MHz) plasma in nitrogen gas (99.99%). The base pressure in the plasma vessel was $10^{-4}$ Pa. The nitrogen flow rate was regulated in the range 80-120 sccm to achieve a constant pressure during implantation of 4.4 x $10^{-2}$ Pa. The rf plasma power was 100 W and the reverse power was 12 W when matched. The plasma density was measured using a Langmuir probe with an rf block (Hiden Analytical Ltd). Acceleration of ions from the plasma was achieved by the application of high voltage (20 kV) bias pulses of 20 μs duration to the sample holder at a frequency of 50 Hz. The plasma treatment system was designed and constructed in house and the PI3 TM high voltage power supply was purchased from ANSTO (Sydney, Australia). A schematic diagram and photograph of the PIII system used is shown in [26].

The samples were mounted on a stainless steel holder, with a stainless steel mesh, electrically connected to the holder, placed 45 mm in front of the sample surface. The samples were treated for durations of 20 - 800 seconds, corresponding to implantation ion fluences of 0.5 - 20 x 1015 ions·cm-2. The ion fluence was calculated from the number of high voltage pulses multiplied by the fluence corresponding to one pulse. The fluence of one high voltage pulse was determined by comparing UV transmission spectra from polyethylene films implanted under the conditions used here to samples implanted with known ion fluences in previous PIII and ion beam treatment experiments [26].

The samples were studied using a visible light ellipsometer in order to estimate the thickness and optical properties (refractive index and extinction coefficient) of the untreated and modified PLGA layers. A Woollam M2000V spectroscopic ellipsometer was used for the measurements. Ellipsometry data were collected for three angles of incidence: $65^0$, $70^0$ and $75^0$ in the wavelength range 400-800 nm. A model consisting of a Cauchy layer on top of a silicon substrate was used to fit the data. Thickness and optical constants associated with the best fit model were determined for the untreated and PIII treated PLGA layers.

Transmission FTIR spectra were recorded using a Digilab FTS7000 FTIR spectrometer and Bomem FTIR spectrometer. FTIR ATR spectra were recorded with a Germanium prism crystal at 45° incident angle. The spectrum of silicon wafer was subtracted to analyze the spectra of PLGA. The optical density of spectral lines associated with particular bond vibrations were used to quantify structural changes in the polymer films.

Microphotos of the polymer surface were taken using a Carl Zeiss Jena microscope with video camera. Objectives of x10, x20 and x50 were used. Digital images were processed with ArtSoft PhotoStudio 4.0 software.

In order to determine the extent of cross-linking (gel-fraction), the ion implanted polymer films were soaked in acetone for 5 days. The samples were then dried in air overnight prior to ellipsometric, FTIR and microscopy characterization.

Simulations of argon ion penetration into a PLGA film were also carried out with the TRIM-95 and SRIM-2003 codes [61, 62] and used to predict the depth-distribution of defects in these films.

For degradation the PLGA samples were soaked in μQ-water water at room temperature (23°C) and at higher temperatures of 37°, 50° and 55°C stabilized using a thermostat with an accuracy of 0.1°C. Given that the glass transition temperature of PLGA is 58°C, the degradation kinetics can be assumed to occur with PLGA in the same physical state [16-18]. The high temperatures were used to accelerate the degradation process. Following van't Hoff law, the rate of degradation at 50° and 55°C increases correspondingly with factor of 2.6 and 3.6 to the rate of degradation at 37°C.

The kinetics of the degradation process were observed by mass measurement, FTIR spectroscopy, ellipsometry and optical microscopy. For mass measurement, the mass of PLGA films on fabric was measured immediately after they were taken out of the solution (wet mass) and the after drying for 2 hours in air and 2 hours in vacuum (dry mass). The swelling was characterized by the ratio of the difference between wet and dry masses to the dry mass of each sample. The swelling of polyester fabric was eliminated. FTIR spectra and ellipsometric data were recorded from the dried PLGA films on silicon wafers to eliminate the water spectral lines. The degradation was observed on absorbance intensity of characteristic lines in FTIR spectra and by thickness measured with ellipsometry.

## Results

### *Plasma immersion ion implantation of PLGA*

The 100 nm PLGA coating on silicon wafer showed no visually detectable changes after PIII-treatment. Under the microscope, the surface of this coating appears smooth (Figure 2). Thick PLGA films (1 mm) subjected to high fluence plasma immersion ion implantation (PIII) become milky white when observed in reflection and light brown in transmission. The film on polyester fabric became light brown after high fluence PIII while it appeared white or transparent after a low fluence treatment.

FTIR transmission spectra of the PLGA coating on silicon pulsed biased at 20 kV as a function of implanted ion fluence change significantly (Figure 3). The intensities of all spectral line of PLGA decrease with ion fluence indicating a reduction in the film thickness. A similar etching process was observed at 10 and 5 keV energies and is commonly observed for polymers subject to plasma discharges [26]. The spectra changes attributed to the structure changes are not visible due to significant decrease of the intensity. To visualize the spectra changes, the same spectra are presented in Figure 4, normalized by 1380 $cm^{-1}$ line intensity. These spectra show structural transformations of the PLGA coating caused by ion implantation. Following TRIM calculations, the nitrogen ions implanted with energy of 20 keV penetrate through the 100 nm PLGA coating and cause structural changes in the entire coating. The IR spectra show that as the fluence increases the lines associated with vibrations in the PLGA macromolecules broaden due to an accumulation of defects. The line associated with the carbonyl vibration at 1750 $cm^{-1}$ develops a shoulder at 1700 $cm^{-1}$ after low fluence of PIII. At high fluence these lines disappear all together remaining a line at 1650 $cm^{-1}$. Lines at 1455-1380 $cm^{-1}$ attributed to C-H vibrations are transformed into one broad band. The same transformations of broadening and overlapping of lines are observed in the 1100-1200 $cm^{-1}$ region of C-O vibrations. Low frequency lines at 881, 901 and 967 $cm^{-1}$ appear in the PLGA spectrum after high fluence ion implantation. These lines are attributed to out-of-plane C-H vibrations in unsaturated carbon-carbon structures. All these spectral changes are indicative of the destroying of the initial PLGA macromolecules and the carbonization occurring in the modified PLGA coating.

FTIR ATR spectra (Figure 5) from a thick PLGA film on fabric for a range of ion implantation fluences show smaller changes than from the thin PLGA film, because the thickness of modified layer, which is about 100 nm, is significantly smaller than the depth of penetration of the infrared probe beam (400-500 nm). The wide band between 1600 and 1700 $cm^{-1}$ appeared with modification corresponds to oxidation and carbonization of the thin (~100nm) surface layer which has been modified.

The refractive index of ion beam treated 100 nm PLGA coating changes dramatically with fluence (Figures 6a and 6b). The values shown were determined using spectroscopic ellipsometry. The refractive index of the unmodified PLGA is in the range of 1.47-1.49 over the 400-1000 nm spectral interval. After ion implantation the refractive index increases to 1.6-1.7. At short wavelengths the refractive index for the modified samples is considerably higher than at low wavelengths that shows chromatic dispersion due to optical absorbance at short wavelengths. This short wavelength absorbance is caused by the appearance of unsaturated carbon structures in the ion modified PLGA. The observed refractive index is consistent with the refractive index of other polymer coatings after ion beam treatment [63-65].

Decreasing thickness of the coating determined from the ellipsometric data is observed with increasing ion fluence due to etching (Figure 7). The etching rate in dependence on ion fluence does not show any significant influence of ions energy. The average etching rate is $(1.46 \pm 0.19) \bullet 10^{-13}$ $nm \bullet cm^2 \bullet ions^{-1}$.

Ion beam implantation creates an insoluble fraction by causing crosslinking of the PLGA macromolecules. To determine the gel-fraction, the modified PLGA coating on silicon were washed in acetone for 5 days. After washing, the residual layer were analyzed by FTIR transmission spectroscopy. No gel-fraction was found for the unmodified sample and for the coating modified at low fluence ($2 \times 10^{14}$ $ions/cm^2$). Figure 8 shows FTIR spectra taken from the insoluble part of the films ion implanted at higher fluence. The spectra contain lines at 3513 $cm^{-1}$ (hydroxyl groups), 2994 $cm^{-1}$ and 2948 $cm^{-1}$ (C-H vibrations in PLGA), 1751 $cm^{-1}$ (carbonyl vibrations of PLGA), 1725 $cm^{-1}$ and 1709 $cm^{-1}$ (carbonyl in ion damaged PLGA), 1650-1600 $cm^{-1}$ (carbonized part of PLGA) and 1455-1380 $cm^{-1}$ (bending C-H vibrations of PLGA). These spectra show that the gel-fraction contains carbonized materials as well as unchanged PLGA macromolecules.

The results presented in this section have shown that ion implantation causes deep structural transformations in the PLGA surface layer, including the carbonization and crosslinking of PLGA macromolecules. While this modification is taking place the PLGA coating is also being etched at a rate that depends on the fluence of the implanting ions.

*Degradation of PLGA film on silicon wafer, steel and carbon layer*

The initial surface of the unmodified spincoated PLGA is smooth (Figure 2). After a 24 hour exposure to water at $50^0$C the coating becomes rough (Figure 9). The surface develops hills and valleys. The valleys become deeper with time and eventually reach the substrate surface. The coating becomes ruptured. After 48 hours the valleys spread until they overlap and separate drops of PLGA appear. During next days the drops remain on the same position and with the same size, but become smaller. After 522 hours the drops are slightly visible spots.

These transformation of the coating can be interpreted as dewetting and degradation processes. At first period of time (48 h) the coating dewets due to instability, when the surface tension forces in the coating are stronger than interface forces between the coating and the silicon wafer [66-75]. There are two kinds of dewetting: dewetting due to defects on the surface of silicon wafer or in the coating and spinodal dewetting. In the first case, the defect initializes the dewetting process, which starts hole formation near the defect. The hole increases in diameter due to the surface tension forces at the edge of the hole. The dewetting process on the defects is connected with the purity of the coating and the substrate. In the case of spinodal dewetting, the coating is ruptured due to coating surface fluctuation amplification in the field of the repulsive interaction between the coating and the silicon wafer. There is characteristic wavelength, which depends on a thickness of coating, its viscosity and coating-substrate interactions. In our case the periodical structure of drops is clear observed. Such cell-like drop distribution corresponds to spinodal dewetting [4]. The diameter of drops observed after 48 h of dewetting is 1-60 μm.

At the second period of time (from 48 to 522 hours) the drops degrades with time. The height of drops goes down. Finally, the drops disappear. This degradation is caused by hydrolysis reaction with decomposition of PLGA macromolecules to lactone and glycol acids monomers and releasing them into water solution [16-18].

The dewetting process depends on temperature of solution. The dewetting is slow at $23^0$C, when the coating becomes rough but remains not ruptured during 6 days (Figure 10). When heating up to $55^0$C is applied, the dewetting proceeds quicker, that is observed by the appearance of separated drops after shorter time (4 hours). The same effect of heating is observed thick coating (1 μm), when dewetting is accelerated with high temperature (Figure 11).

The PLGA coating dewetting is observed on a stainless steel surface (Figure 12). The dewetting of PLGA on stainless steel is quicker than on the silicon wafer. After 2 hours the coating is roughened and becomes ruptured after 4 hours of exposure. This is explained by higher difference of the surface energies of PLGA and metal, than PLGA and silicon. However, the spinodal dewetting is not observed due to high roughness of the stainless steel surface. The dewetting of PLGA on metal proceeds on topological defects of the metal surface due to high sensitivity of the dewetting to topography of the surface as it was observed for polystyrene coating [74-78].

The dewetting of PLGA was observed on carbonized coating. For this experiment the carbonized coating was prepared on the silicon wafer by the technology, described in [64]: polystyrene coating was modified by ion beam up to complete carbonization. Then PLGA coating was spun on the carbonized surface. In water the dewetting of PLGA coating proceeds on topological defects of the carbon coating (Figure 13). Non-spinodal dewetting is observed.

The dewetting is observed in FTIR transmission spectra. The spectra of the smooth initial coating on carbonized layer show the vibrational lines PLGA (Figure 14). After a 24 hour incubation of the coating in water, the intensity of spectra goes down and the lines are accompanied negative satellites, which correspond to a combination of the absorbance and reflectance spectra of PLGA. The coating is transformed into separate drops. The infrared beam of spectrometer passes partially through the drops giving absorbance lines, partially reflects on drop's surface and partially passes through the naked substrate without absorption. As consequence, the spectrum has less contrast and the lines are deformed with negative satellite. After incubation of the coating for 48 hours and longer, the lines of spectra do not change, but the intensity of all lines goes down. This period corresponds to the hydrolysis of the PLGA drops. The spectra do not show any changes of the chemical structure of the PLGA, implying a uniform degradation and releasing for the lactone and glycol parts of PLGA macromolecule. The released monomers do not remain in the drops and spectral lines of monomers are not presented.

The dewetted PLGA coating could be dangerous for blood stream, when low diameter vessels could be blocked by peeled off drops of PLGA. The size of observed PLGA drops is between 1 and 60 μm. This is equal or bigger than the capillary vessel size (5-10 μm). The intensive blood stream and moving of the stent can assist the PLGA drops to peel off. Therefore, the dewetting of the PLGA coating should be stopped.

*Degradation of PLGA film after ion implantation*

The degradation process of PIII modified PLGA coating is different than in untreated PLGA coating. Figures 15-19 show images of PLGA coatings of 1 μm thickness on silicon wafer after different fluence of PIII treatment and exposition to water at 50ºC.

At low PIII fluences as shown in Figures 15 ($2\times10^{14}$ ions/cm$^2$) the gel-fraction determined after washing of PLGA coating in acetone is not observed. After PIII treatment with fluence $10^{15}$ ions/cm$^2$ (Figure 16) and higher (Figure 17-19) the gel fraction remains well visible on the silicon wafer as cracked film.

At low PIII fluences ($2\times10^{14}$ and $10^{15}$ ions/cm$^2$) the dewetting of the coating is similar to that of the untreated PLGA film in that the coating is roughed and ruptured at the beginning and followed by the formation of separate drops. In contrast to the behaviour of the untreated films, the drops are distributed irregularly and most of the drops do not have symmetrical shapes. It means, that the dewetting is not spinodal and does not proceed on the defects of the silicon wafer. The modified surface also contains sections where the dewetting process stops at the ruptured stage. In such areas, the valleys form as for the untreated PLGA but the drop formation is not observed and the PLGA films remains in tact in water for all subsequent observation times. In the final stage of degradation, the height of the drops decreases. The part of the PLGA film which has not dewetted remains in place while its height also decreases with time in water.

At higher fluences (from $2\times10^{15}$ to $2\times10^{16}$ ions/cm$^2$), the part of PLGA coating dewets to up to separate drops as described previously (Figure 17-19). But other part of PLGA does not dewet upon exposure to water. The not dewetted sections of PLGA cover about half of surface. Some drops are also observed on the non-dewetted areas. The remained film is viewed shrunk and cracked. The color of the remained film referred its thickness becomes darker with fluence of PIII treatment. In general, the remained film view corresponds to the view of gel-fraction of PLGA coating after washing in acetone. The drops of the dewetted part degrade with time in water. The non-dewetted parts of PLGA coating do not disappear after long time in water (522 hours at $50^0$C).

The dewetting and disappearance of the drops does not depend on fluence of PIII treatment. The coating becomes rough after 24 hours and s dewets after 48 hours in water. The drops become non-visible after 522 hours in water for all used fluences of PIII treatment.

The FTIR spectra show differences in the degradation and dewetting processes for PIII treated PLGA coatings. The difference is observed at short time of exposure in water (24 hours at $50^0$C). The negative satellite of the carbonyl line is weaker in the spectra of PIII treated coatings, than in untreated coating (Figure 20). After longer time 72 h exposure in water (Figure 21) the spectra of PIII modified PLGA shows an additional line at 2924 cm$^{-1}$ and the intensity of the lines at 2880 cm$^{-1}$ and 2853 cm$^{-1}$ is higher that corresponds to vibrations of additional CH and $CH_2$ groups of the crosslinked PLGA film in comparison to untreated coating. These spectra differences correspond to non-dewetted part of PLGA coating.

The difference in degradation process is observed in FTIR spectra of degradation products of untreated and PIII treated PLGA coatings (Figure 22). For these spectra the water solution after degradation of the PLGA coating was dried on ATR Germanium crystal. After complete drying, the spectra of residuals on ATR crystal were recorded. The most spectra are similar. But the intensity of the 1750 cm$^{-1}$ line from the carbonyl group of PLGA and a shoulder at 1660 cm$^{-1}$ line of acid carbonyl is higher for the modified PLGA film products, than for the unmodified PLGA. Similarly, the intensity of 1453 cm$^{-1}$ line is higher for the PIII modified PLGA. These lines show, that the products of degradation from the PIII treated coating contain not only lactone and glycol acids monomers, but the products of crosslinking also. These products are associated with the crosslinked fractions of the PLGA molecules, which are not degradable.

The swelling of PLGA in water is a first effect in the dewetting and degradation processes. The swelling kinetics of PLGA in water at $50^0$C was measured for untreated and PIII treated PLGA films of 15 μm thickness on polyester fabric (Figure 23). These films were PIII treated on both sides. The measured curves of the swelling have an increasing part due to water adsorption and a decreasing part due to degradation of the PLGA film. The observed differences in swelling kinetics of untreated and treated films with various PIII fluences are not more than the experimental error.

The mass loss measurements show similar kinetics of degradation for untreated PLGA and PIII treated PLGA films on polyester fabric (Figure 24). The mass loss occurs due to the hydrolysis reaction by which PLGA degrades. Our measured curve corresponds well to reference data of PLGA degradation [9, 10].

All these results of degradation kinetics are related to PLGA films with thickness bigger than the ion penetration depth at PIII. In the case of nitrogen ions of 20 keV energy the modified layer is expected no more than 100 nm. At lower energy of ions, the thickness of modified layer is less. Therefore, all these results are related to the thick PLGA coating with thin modified surface layer.

The investigation of the degradation process in modified layer was done for the thin PLGA coatings of 100 nm which were implanted with nitrogen ions of energy 20, 10 and 5 keV. At 20 keV, the ions penetrate through the coating reaching the silicon substrate. At 10 and 5 keV the ions penetrate about one half and a quarter of the coating's thickness, respectively. Microphotos of the ion implanted PLGA films after incubation in water at 50ºC are presented in Figure 25. Over 4 days of degradation the dewetting effect was not observed for any of the modified PLGA films. On day 4, black traces of PLGA gel-fraction, which have not degraded completely, are observed for 10 and 5 keV energy implanted coatings. The surface remains smooth during the degradation. After 4 days, some amount of PLGA coating remains on the silicon wafer.

The modified PLGA coating has not degraded completely after a long exposure time in water. The degraded thickness of PLGA coating after 4 days in water was measured by ellipsometry for different fluences and energies of ions at PIII (Figure 26). The untreated PLGA coating degraded completely after 4 days in water. The degradable fraction of PIII treated PLGA coating decreases with increasing fluence. At low fluences ($10^{14}$ ions/cm$^2$) the amount of degradable PLGA is about 80% of the initial film. Thus, indicating that low fluence of PIII treatment with high energy provides absence of the dewetting and 80% of degradation of the PLGA coating. The residual part of the coating is not degradable due to crosslinking and carbonization.

*PLGA coating on stents and biodegradation after ion implantation*

The dewetting effect described above is also observed during degradation on PLGA coated vascular stents (Figure 27). The initial PLGA coating on the stents is smooth. After PIII treatment by nitrogen ions with energy of 20 keV and fluence of $5\times10^{15}$ ions/cm$^2$ the coating remains smooth as initial.

After 10 days of incubation in water at 50ºC the untreated PLGA coating is ruptured and dewetted. The stent surface becomes to be coated by drops of PLGA. A part of bare metal surface of stent becomes uncoated. The dewetting is non-spinodal due to curvature of the stent surface.

On the PIII treated stents the PLGA coating incubated 10 days in water at 50ºC appears as a wrinkled film covering the entire surface of the stent. The opened bare metal surface is not observed. The dewetting is stopped.

## Conclusion

The results of our investigation of ion implantation of PLGA show that the PLGA films undergo crosslinking, carbonization and etching. The PLGA coating degrades in water by a hydrolysis reaction. The dewetting of PLGA coating under water is observed on silicon wafer, metal and carbon surfaces. As the result, the coating is ruptured up to separate drops of PLGA on the surface. The dewetting process can be eliminated by an ion beam implantation treatment, when the ions penetrate through the whole PLGA coating. The dewetting process in the modified layer can be eliminated even with low fluence ($10^{14}$ ions/cm$^2$) modification. The amount of degradable PLGA in the modified layer depends on the fluence of ion implantation and can equal up to 80% of the film at low fluence of $10^{14}$ ions/cm$^2$.

## Acknowledgments

The study was supported by Boston Scientific. Authors thank Dr. Jan for support of these studies. Authors thank Mr. Stacey Hirsh for useful discussion and manuscript corrections.

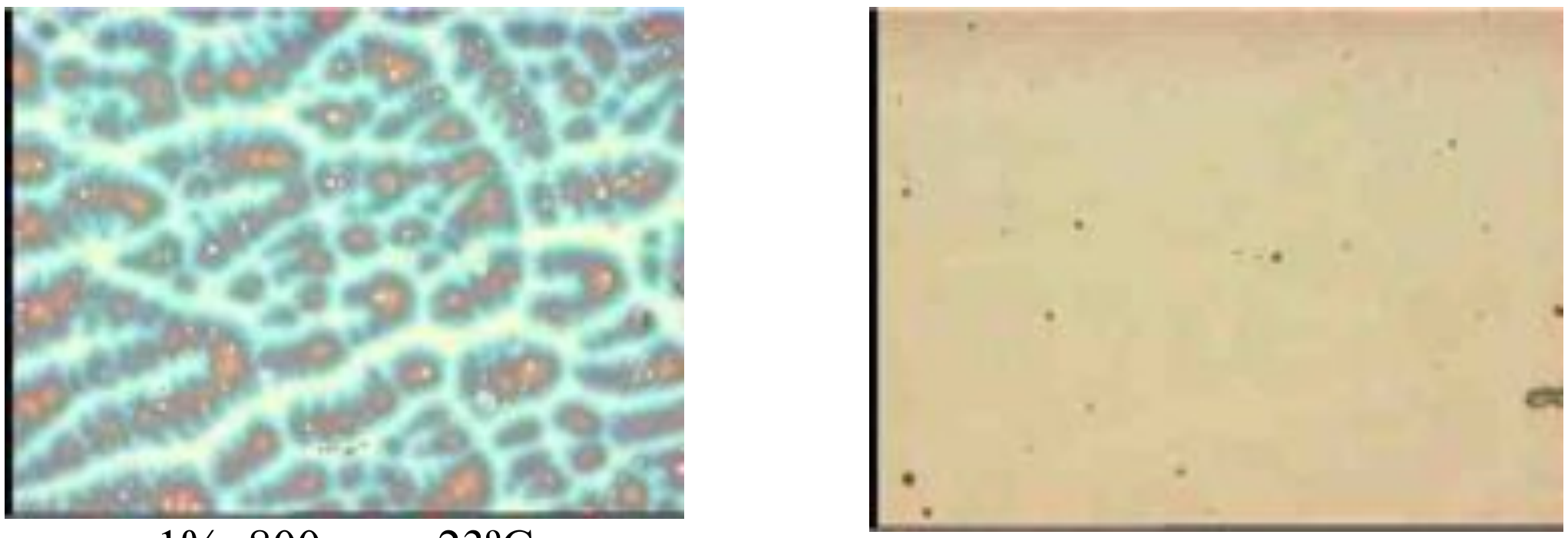

1%, 800 rpm, 23ºC

10%, 1500 rpm, 35ºC, 1 μm

Fig.1. Microphotos of PLGA spun on silicon wafer under different conditions. The microstructure of the film reflects phase separation at low temperature of spincoating. At high temperature, the microstructure (phase separation) is not observed.

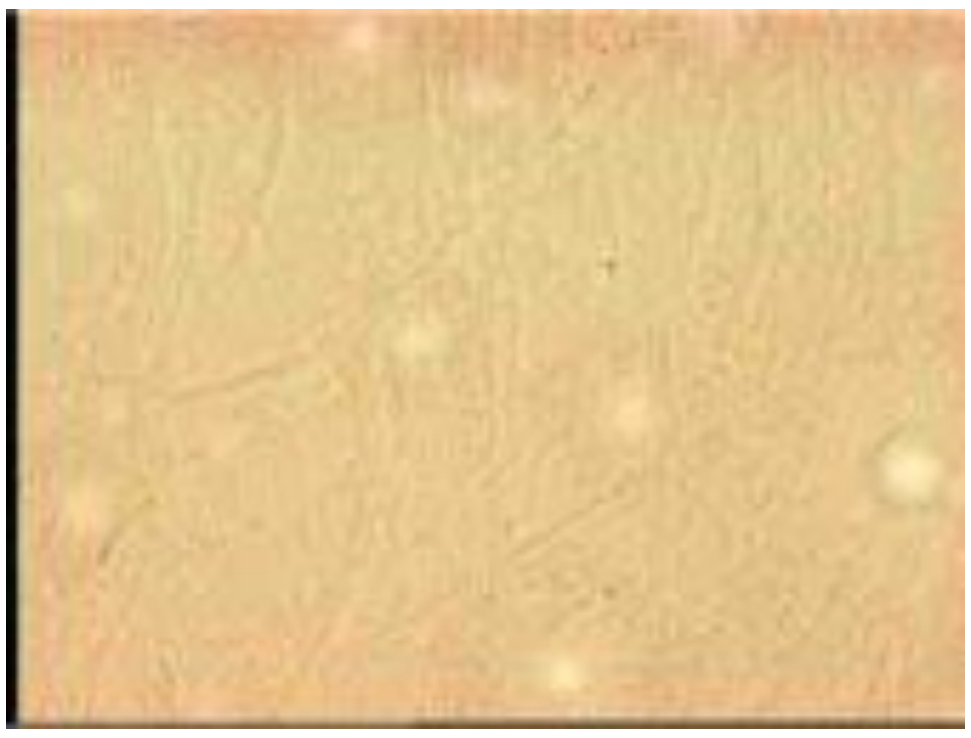

Fig.2. Microphotos of PLGA coating on silicon wafer after PIII with nitrogen ions of 20 keV energy and $5\times10^{14}$ ions/cm$^2$ fluence. Size of photos is 1200x950 μm. The topography of the surface does not change with PIII treatment.

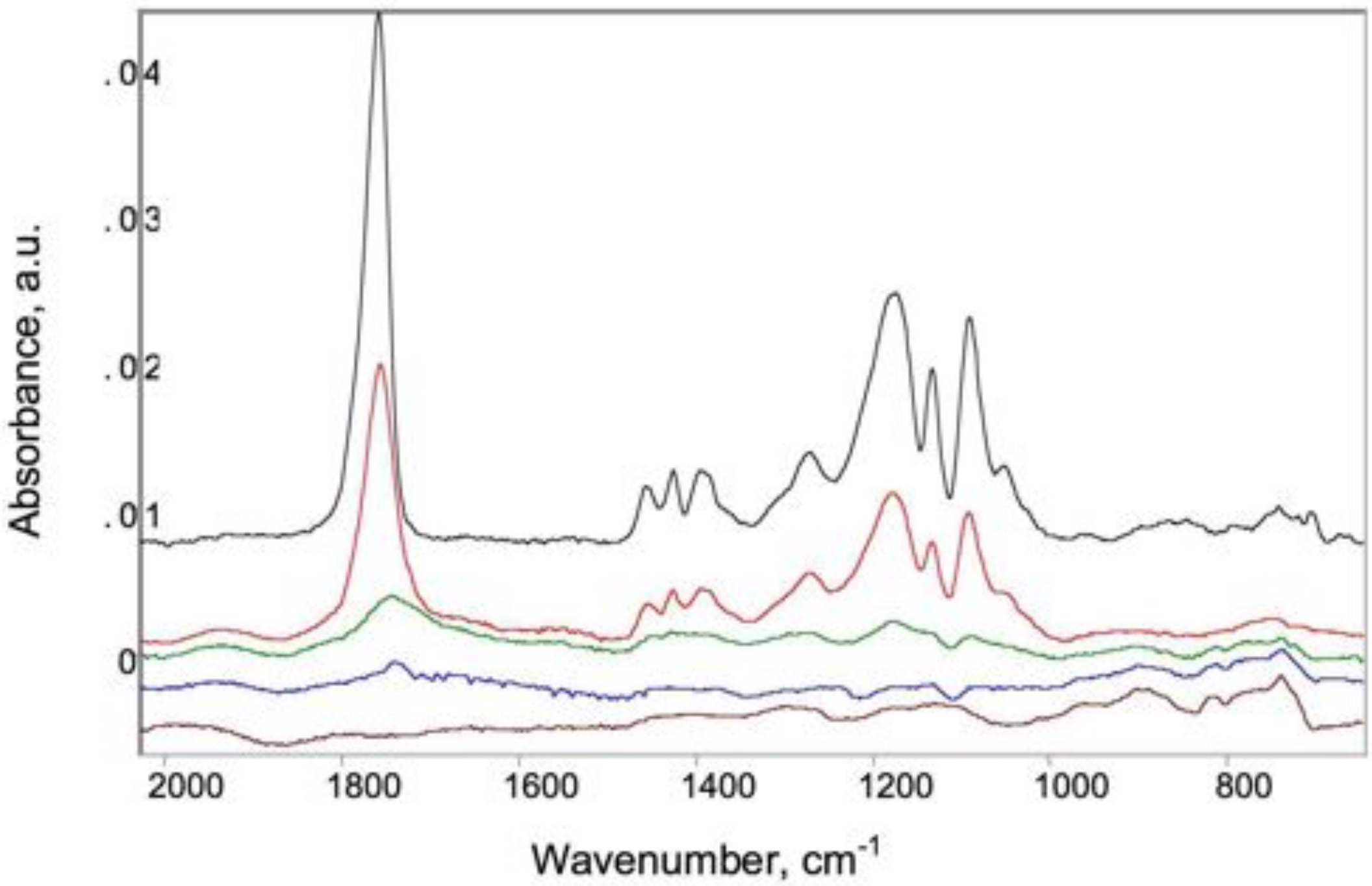

Fig.3. FTIR transmission spectra of PLGA on silicon wafer after PIII with 20 keV ion energy: black – initial, red – $10^{14}$, $5\times10^{14}$, $10^{15}$, $2\times10^{15}$ ions/$cm^2$. Spectra are off-set for viewing. Spectra of silicon wafer are subtracted. The absorbance decreases corresponding to the thickness decrease due to etching of the coating under PIII.

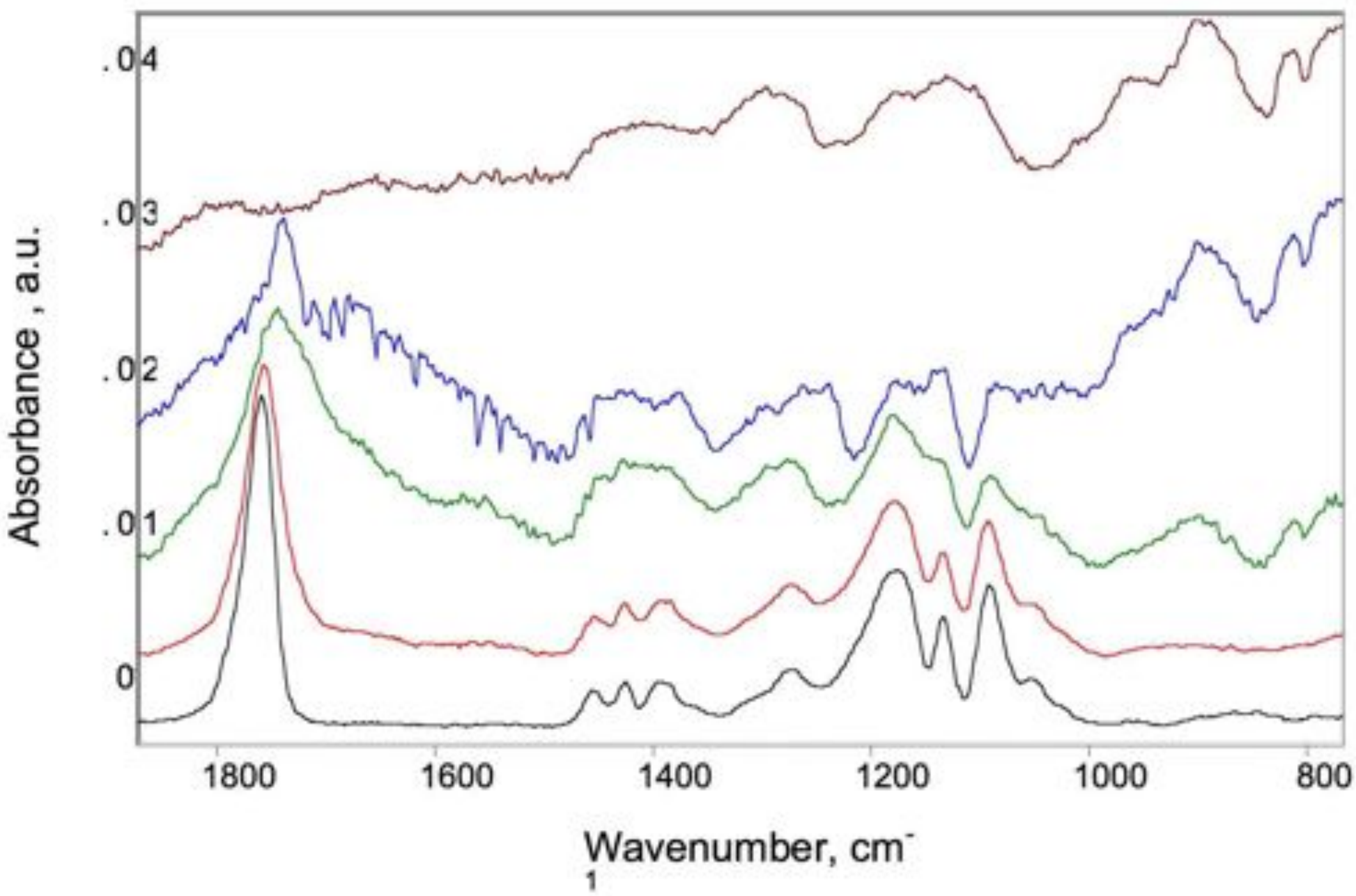

Fig.4. FTIR transmission spectra of PLGA on silicon wafer after PIII with 20 keV ion energy: black – initial, red – $10^{14}$, $5\times10^{14}$, $10^{15}$, $2\times10^{15}$ ions/$cm^2$. The spectra have been normalized by 1380 $cm^{-1}$ line. Spectra of silicon wafer are subtracted. The spectra show chemical transformation of PLGA. Interpretation is in the text.

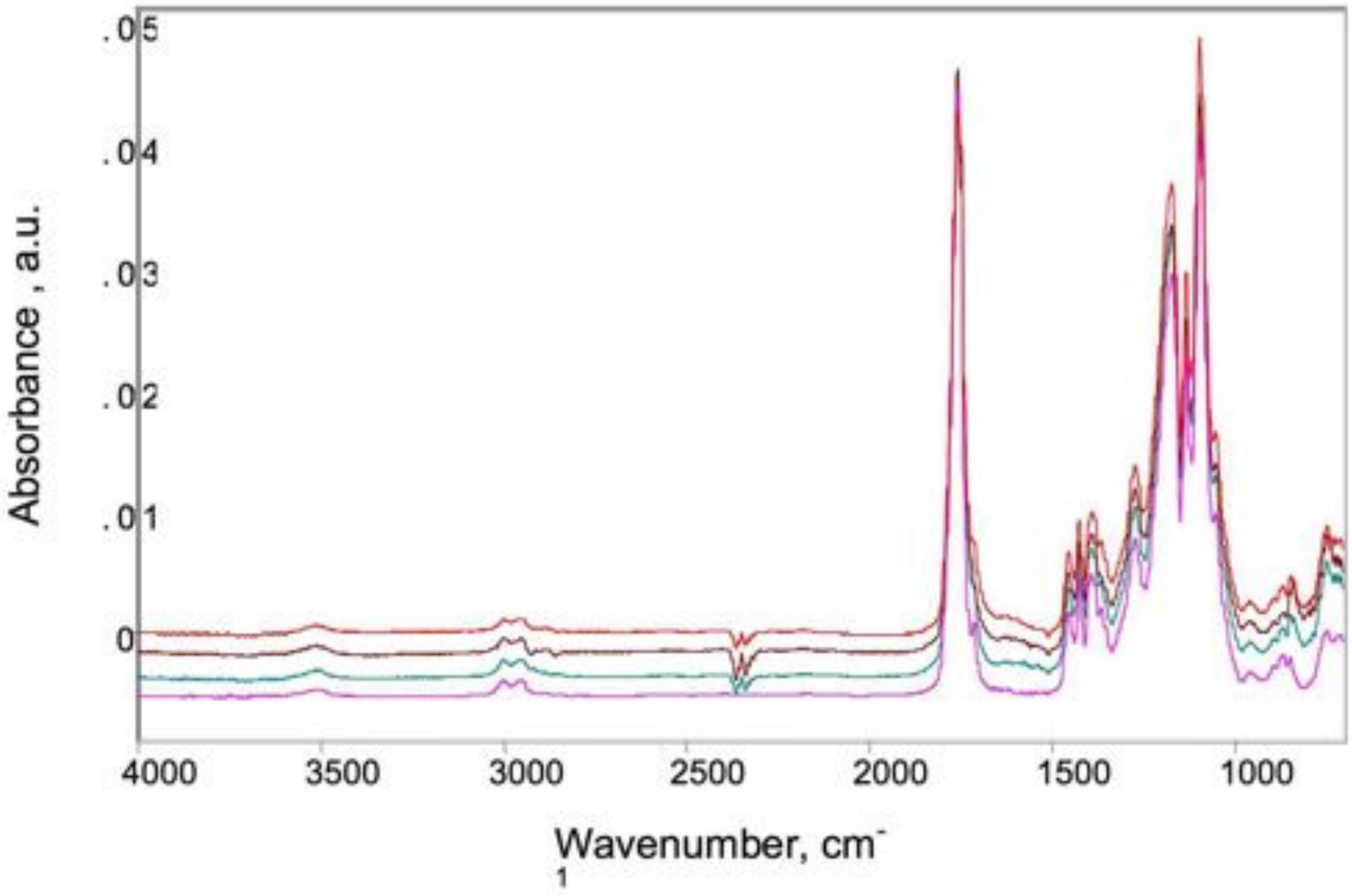

Fig.5. FTIR ATR spectra of PLGA on polyester fabric: pink – initial, red – $5\times10^{14}$, brown – $5\times10^{15}$, blue – $2\times10^{16}$ sec of PIII fluence. The changes are minor. Low intensity band becomes visible at 1650 $cm^{-1}$ in the films after PIII.

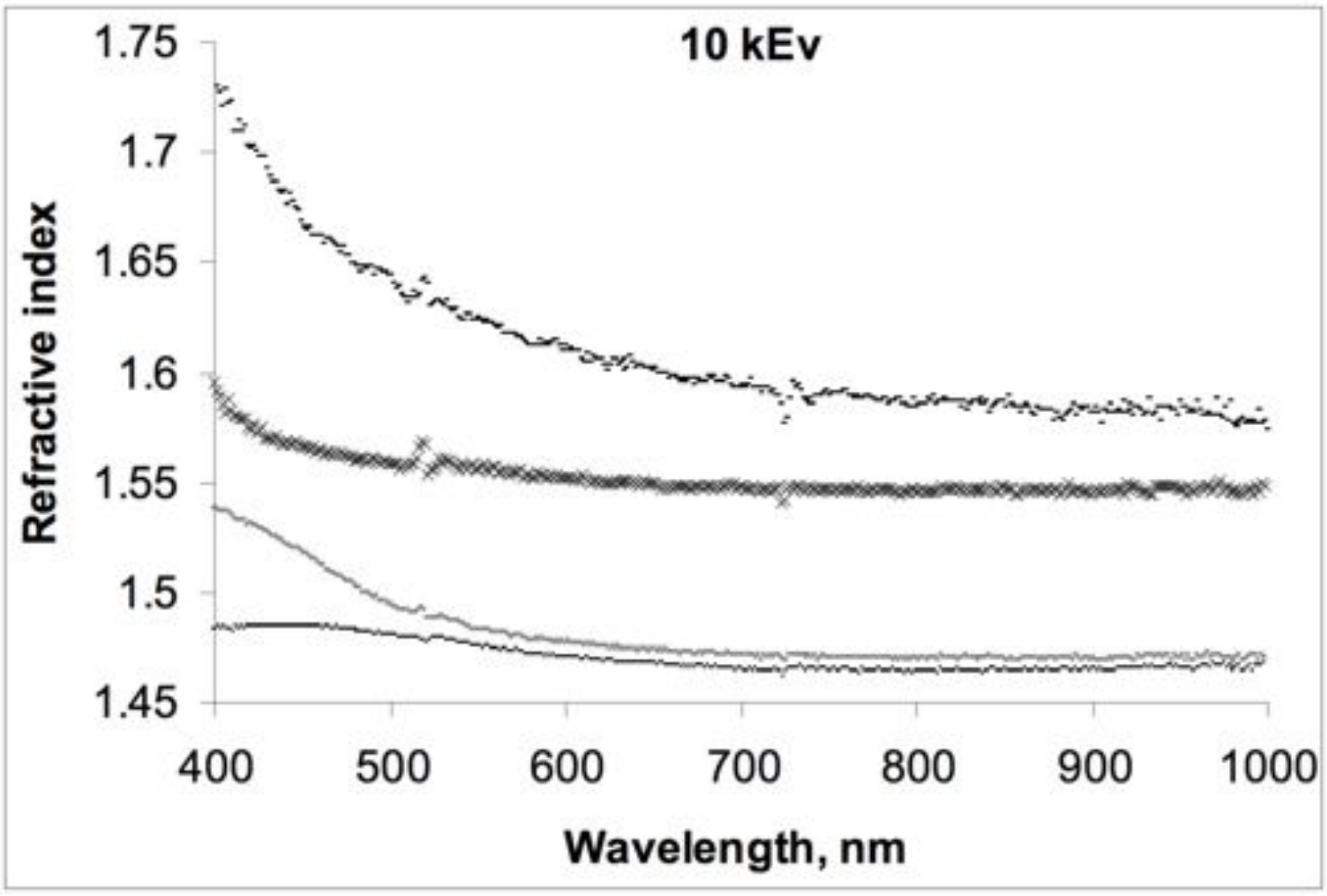

Fig.6a. Refractive index of a 100 nm PLGA layer after 10 keV PIII. From bottom: 0, $10^{15}$, $5\times10^{15}$ and $10^{16}$ ions/cm$^2$ fluence.

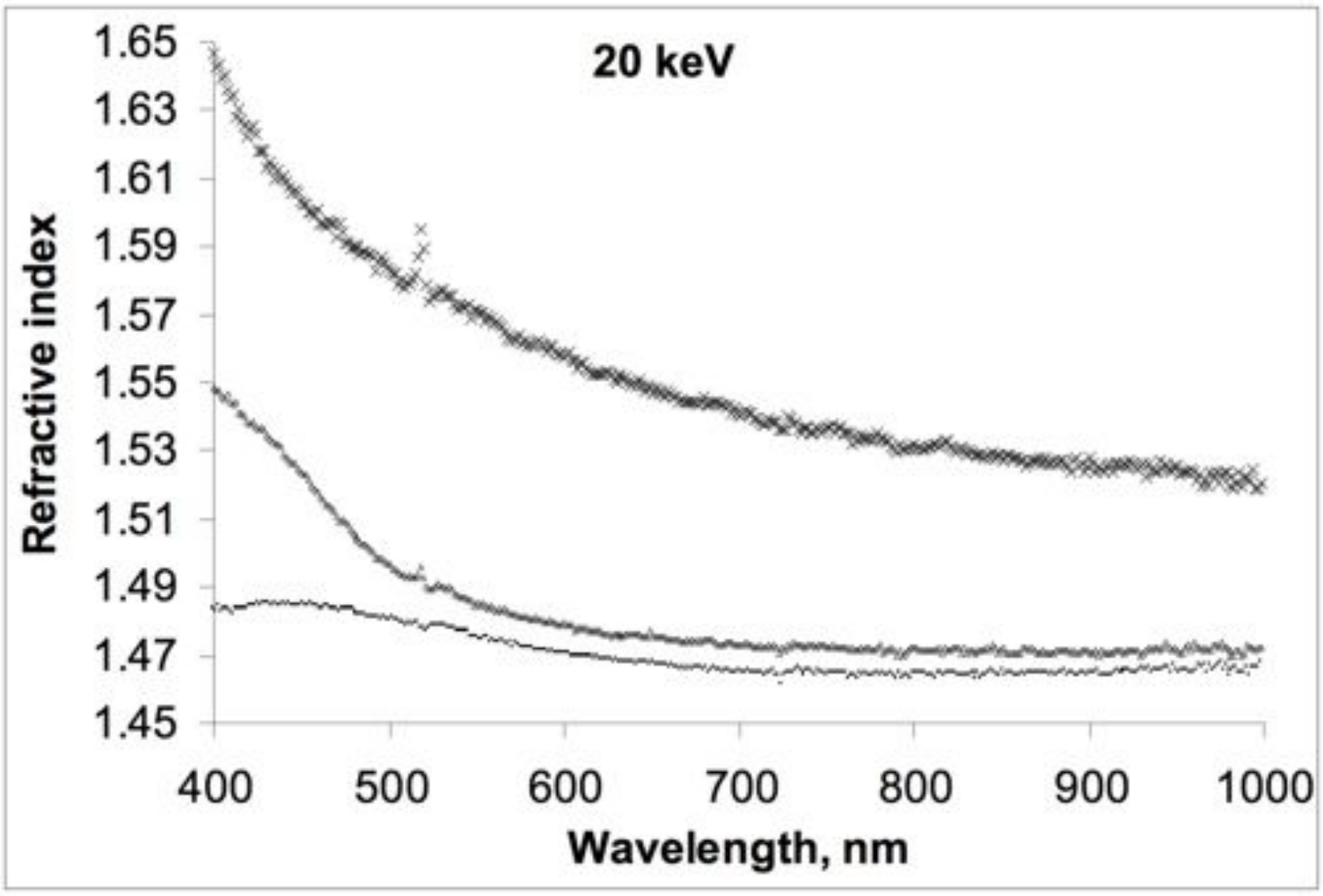

Fig.6b. Refractive index of a 100 nm PLGA layer after 20 keV PIII. From bottom: 0, $10^{15}$ and $5\times10^{15}$ ions/cm$^2$ fluence.

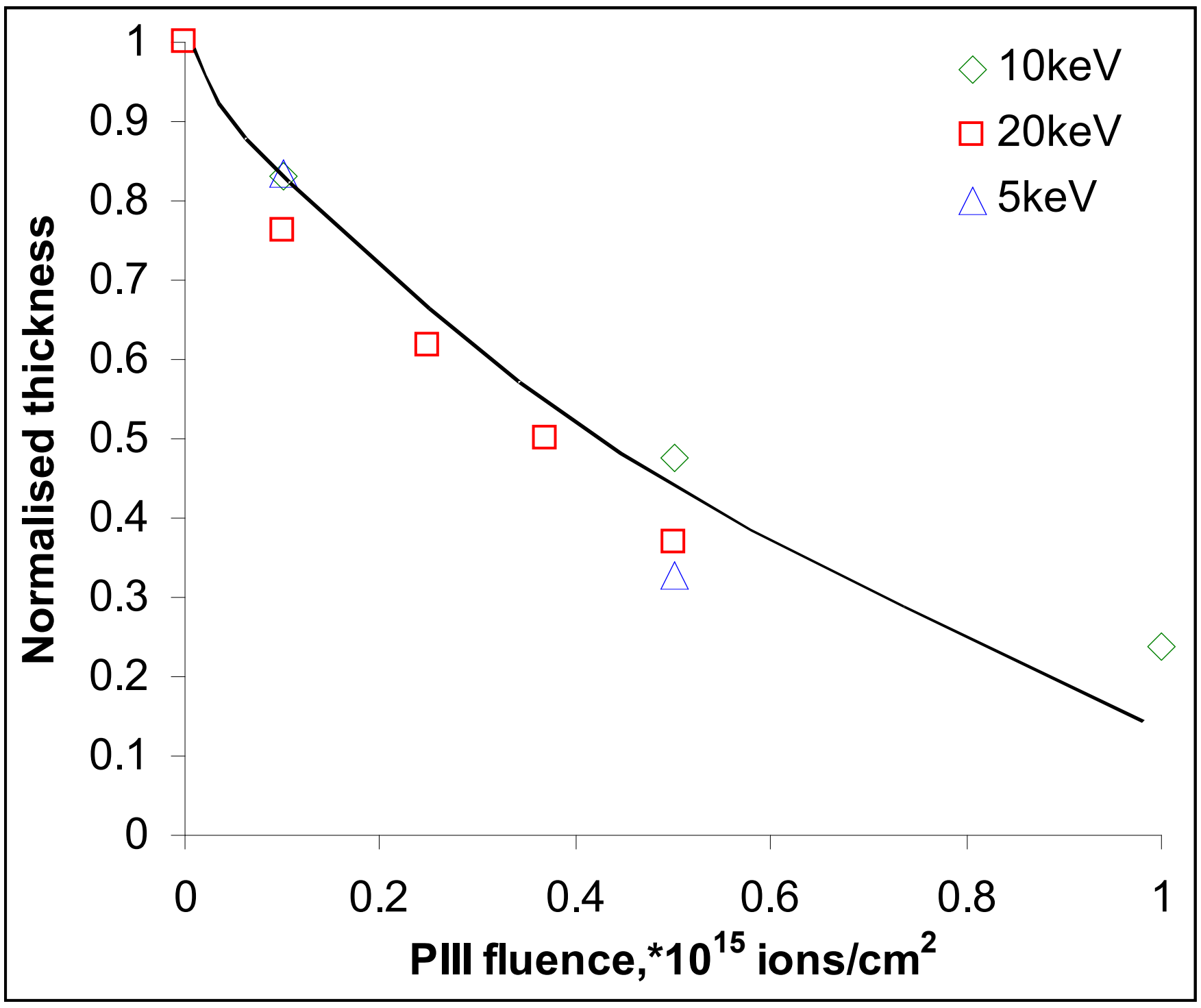

Fig.7. Normalized thickness of PLGA coating in dependence on PIII fluence at different energies of ions.

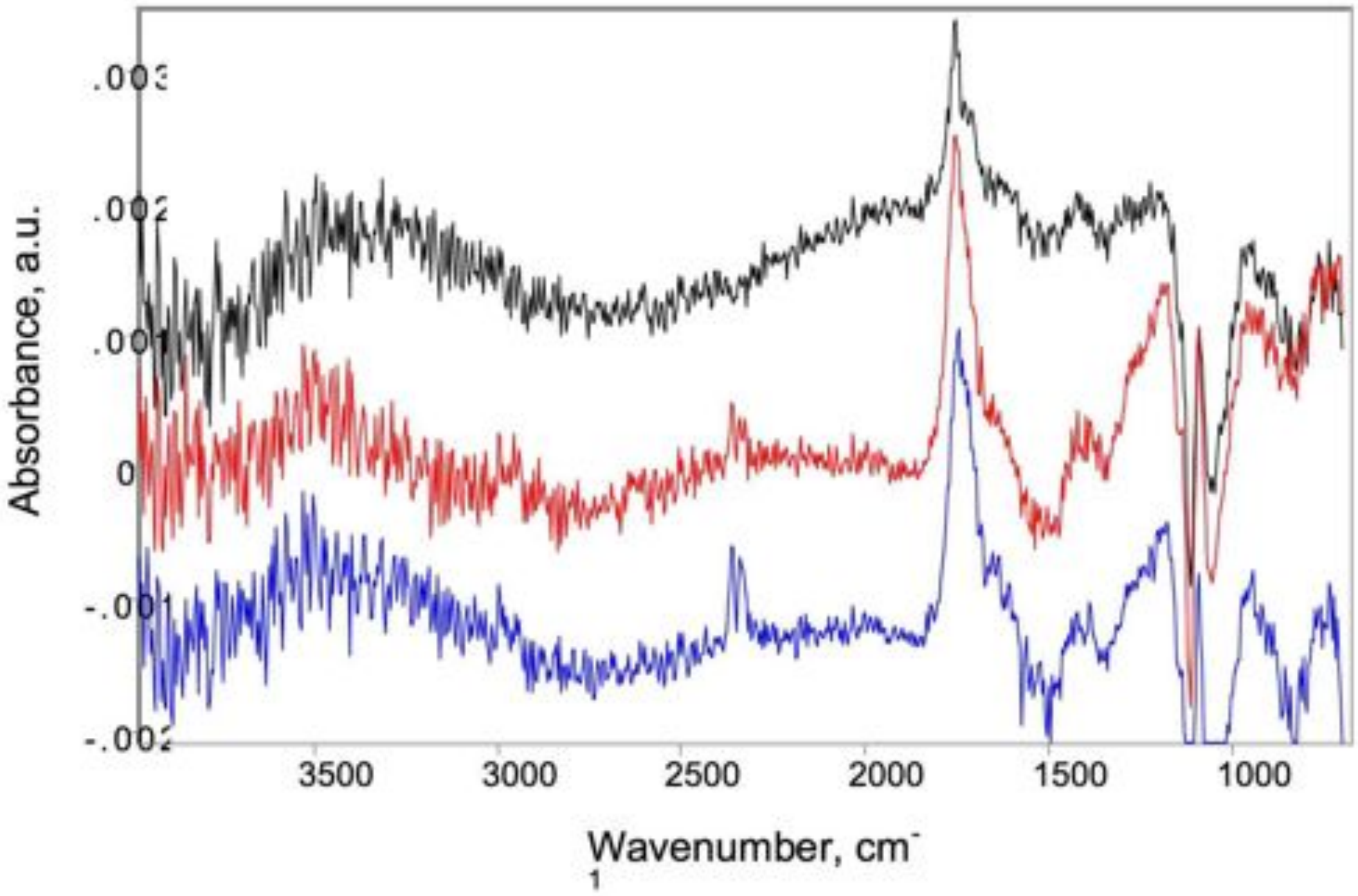

Fig.8. FTIR spectra of gel-fraction of PLGA after PIII and soaking in acetone for 5 days. From the bottom the PIII fluences used were: $10^{15}$, $2\times10^{15}$, $10^{16}$ ions/cm$^2$.

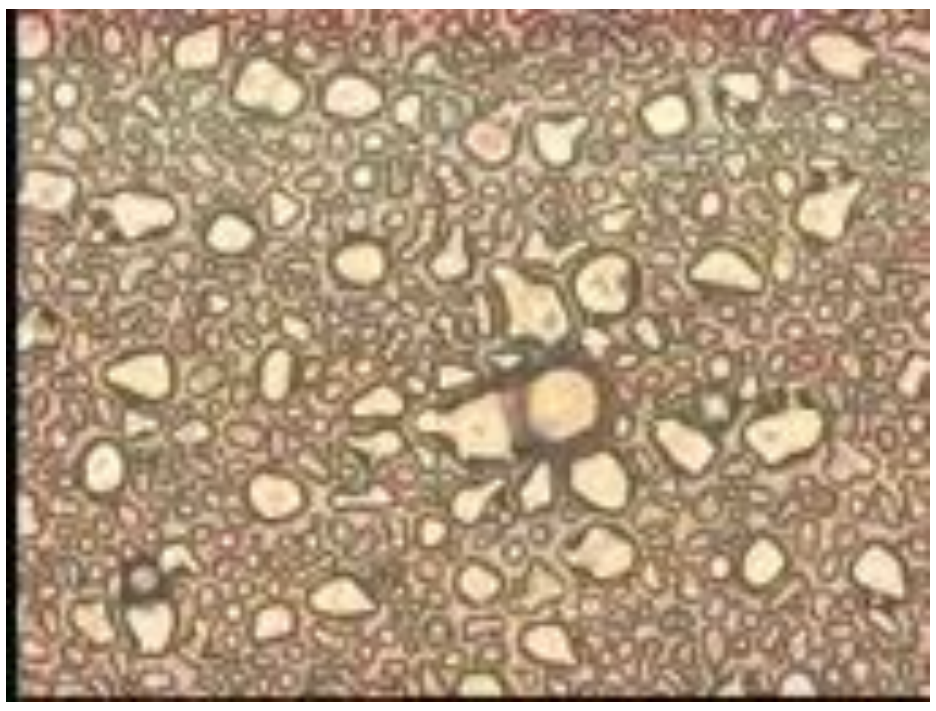

After 24 h in 50ºC water

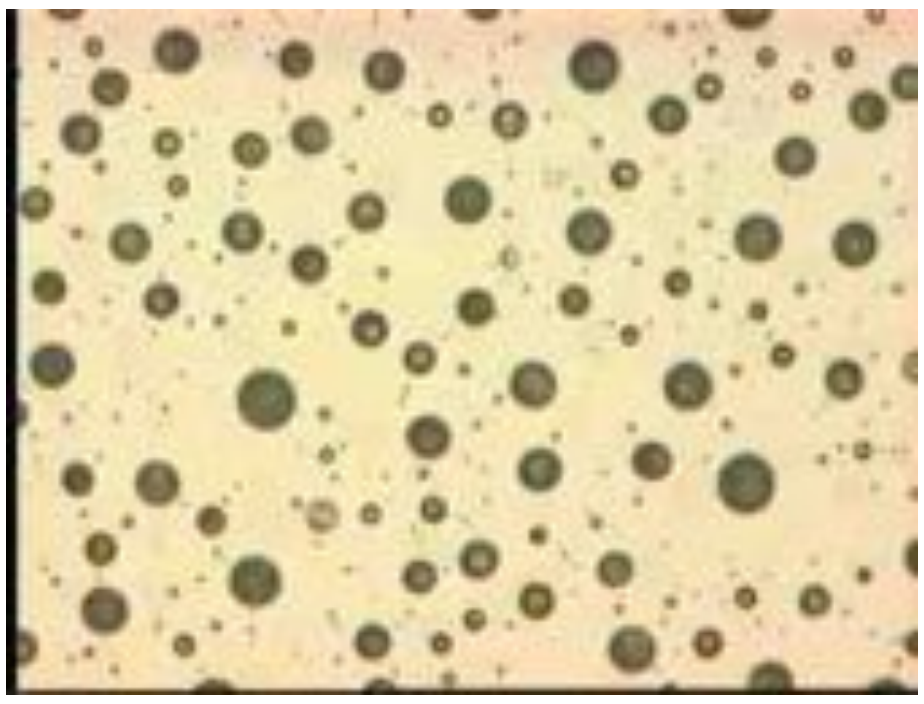

After 48 h in 50ºC water

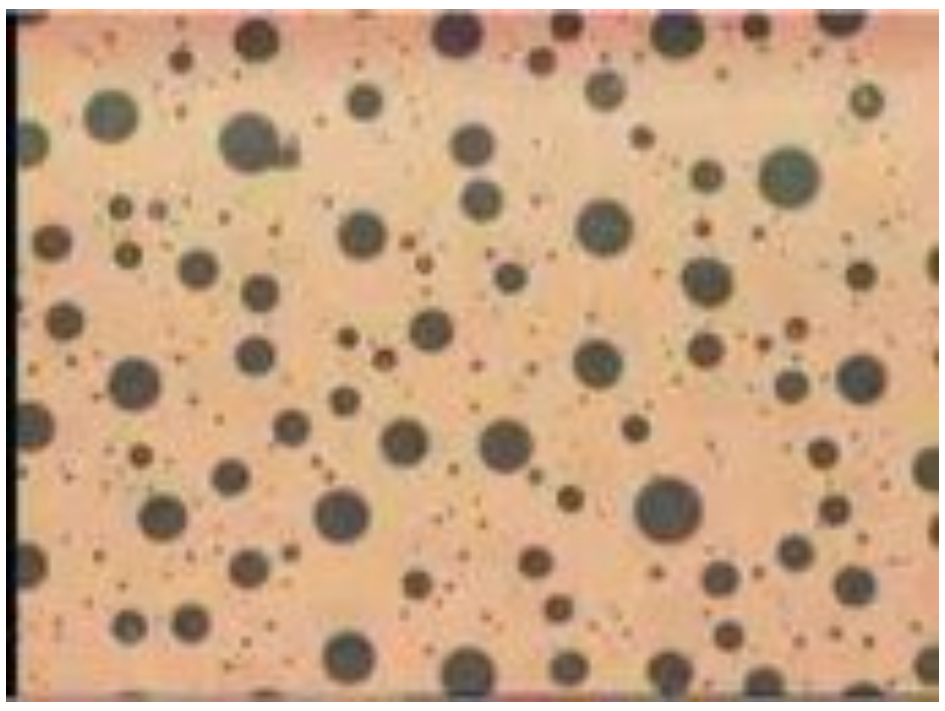

After 72 h in 50ºC water

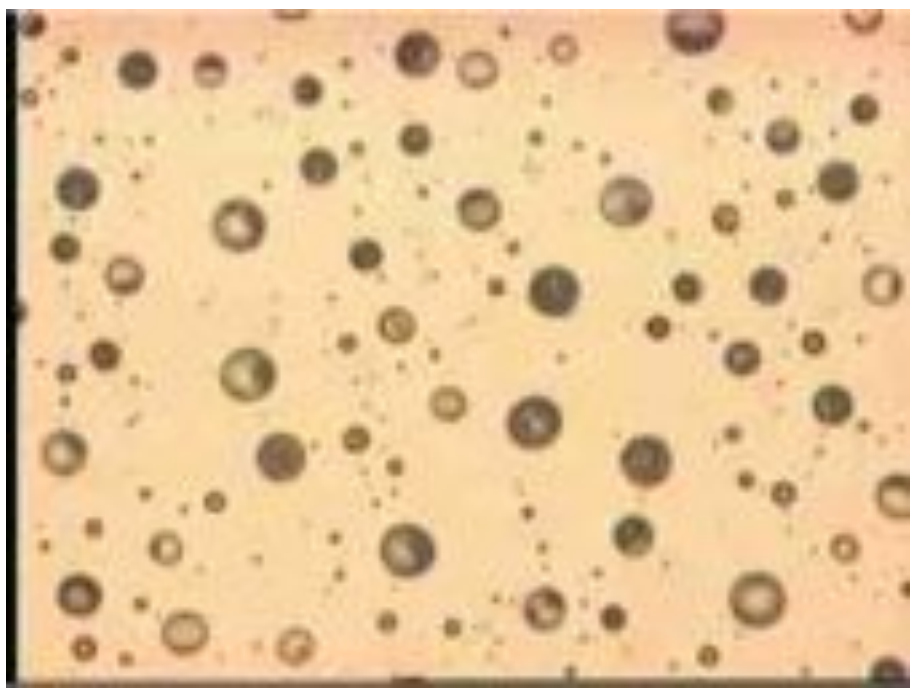

After 120 h in 50ºC water

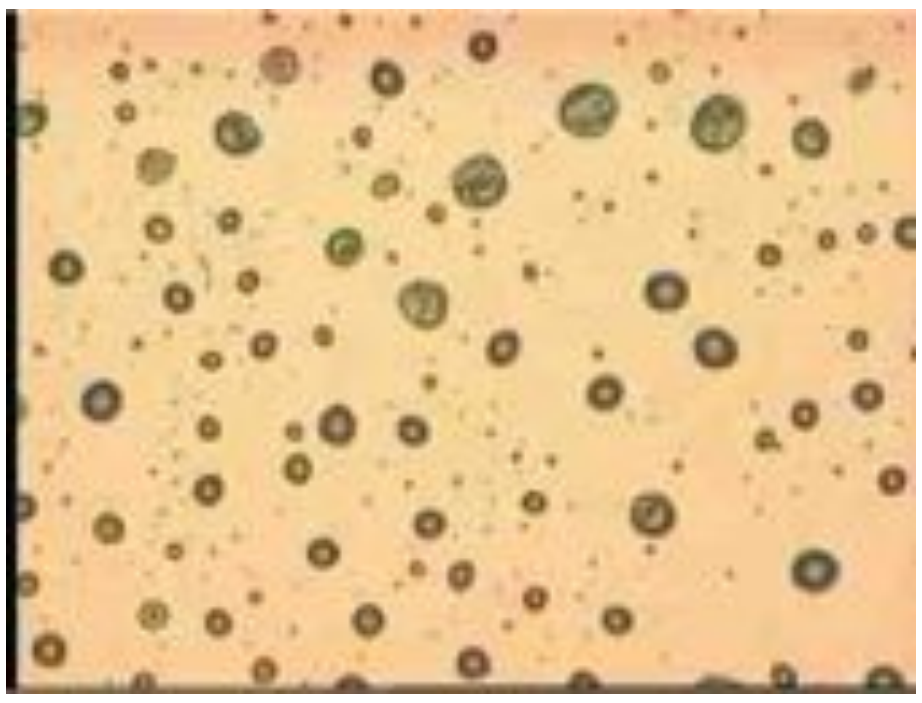

After 216 h in 50ºC water

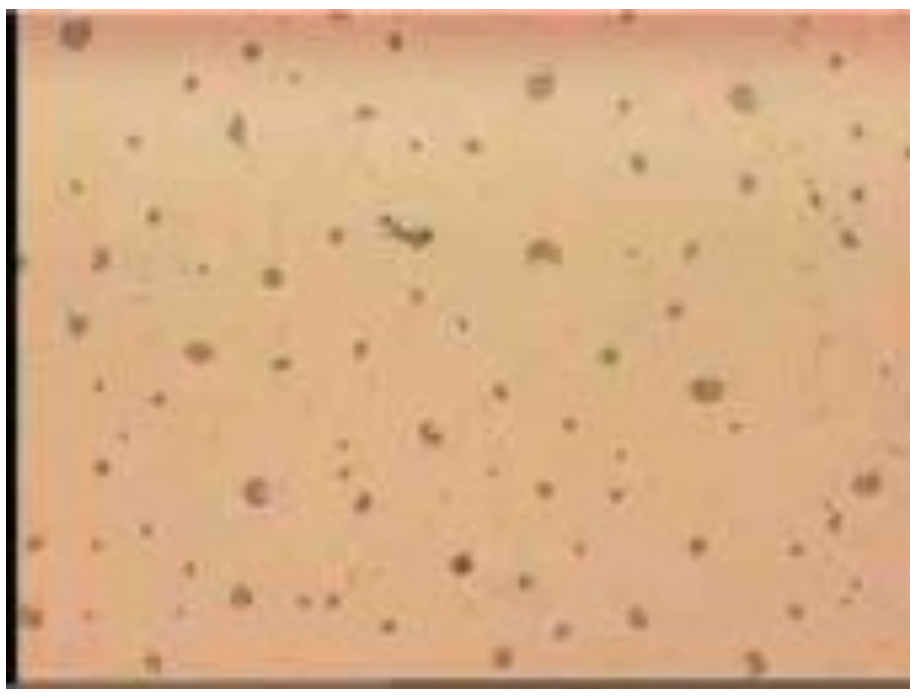

After 522 h in 50ºC water

Fig.9. Microphotos of 1 μm untreated PLGA on silicon wafer after exposure in water. Size of photos is 1200x950 μm.

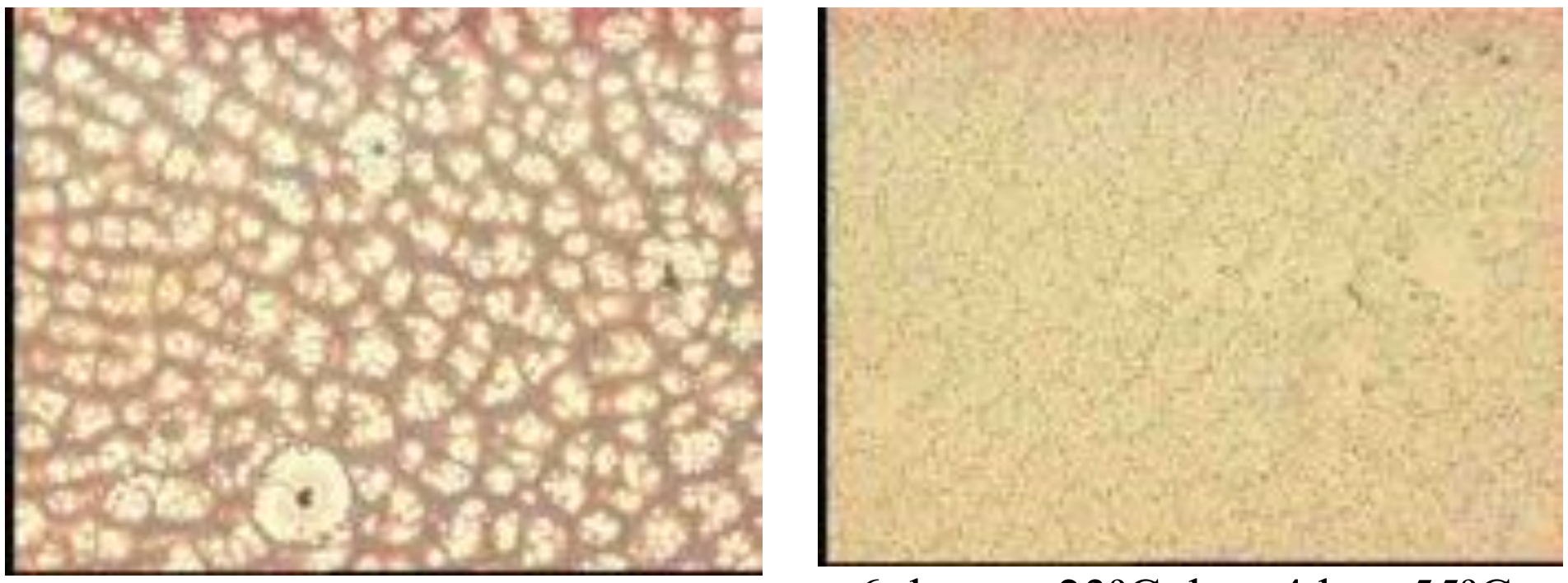

3 days at 23ºC, 1200x950 μm

6 days at 23ºC then 4 h at 55ºC, 1200x950 μm

Fig.10. Microphotos of 100 nm thick PLGA films on silicon after exposure to water. The number of days of exposure and temperature is marked at the base of each photo.

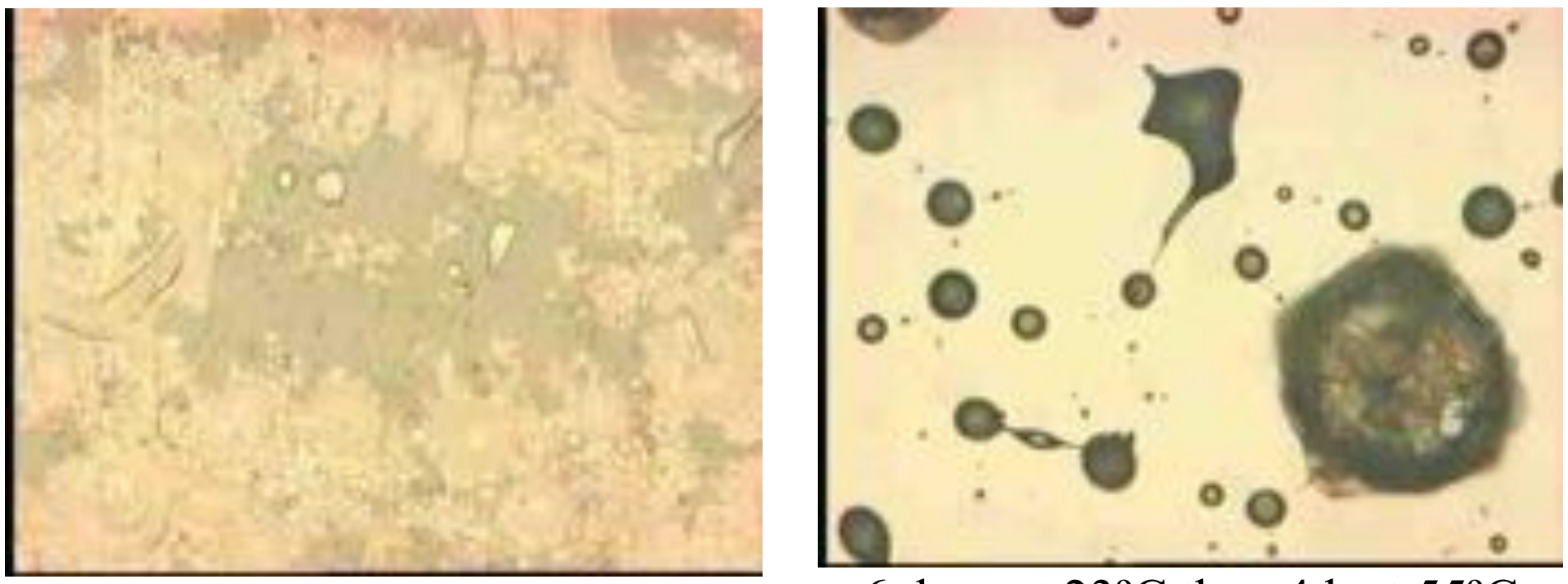

3 days at 23ºC

6 days at 23ºC then 4 h at 55ºC

Fig.11. Microphotographs of 1 μm thick PLGA films on silicon after incubation in water. Size of photos is 1200x950 μm.

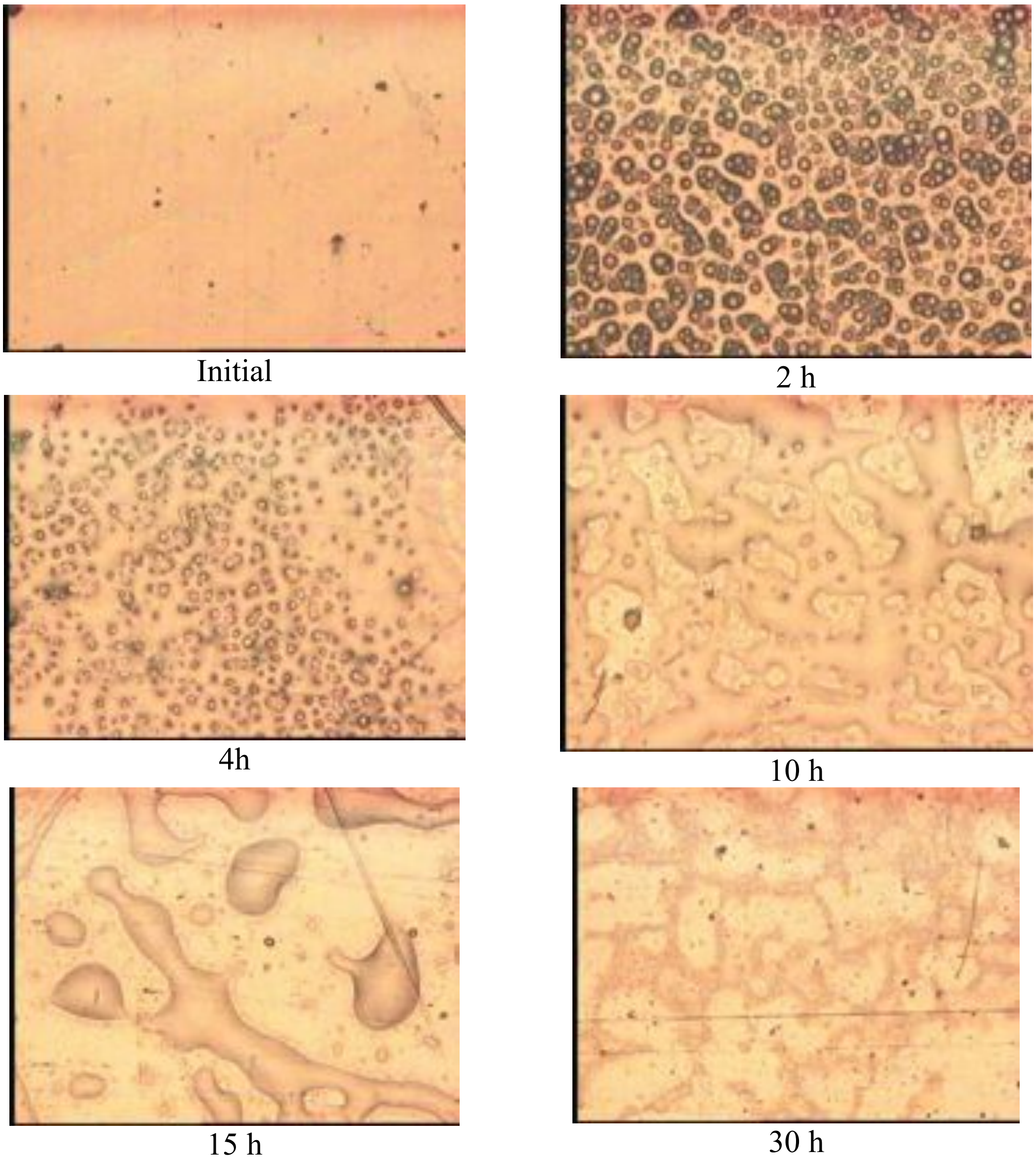

Fig.12. PLGA coating of 1 μm thickness on steel plate after water exposure at 55ºC. Size of photos is 1200x950 μm.

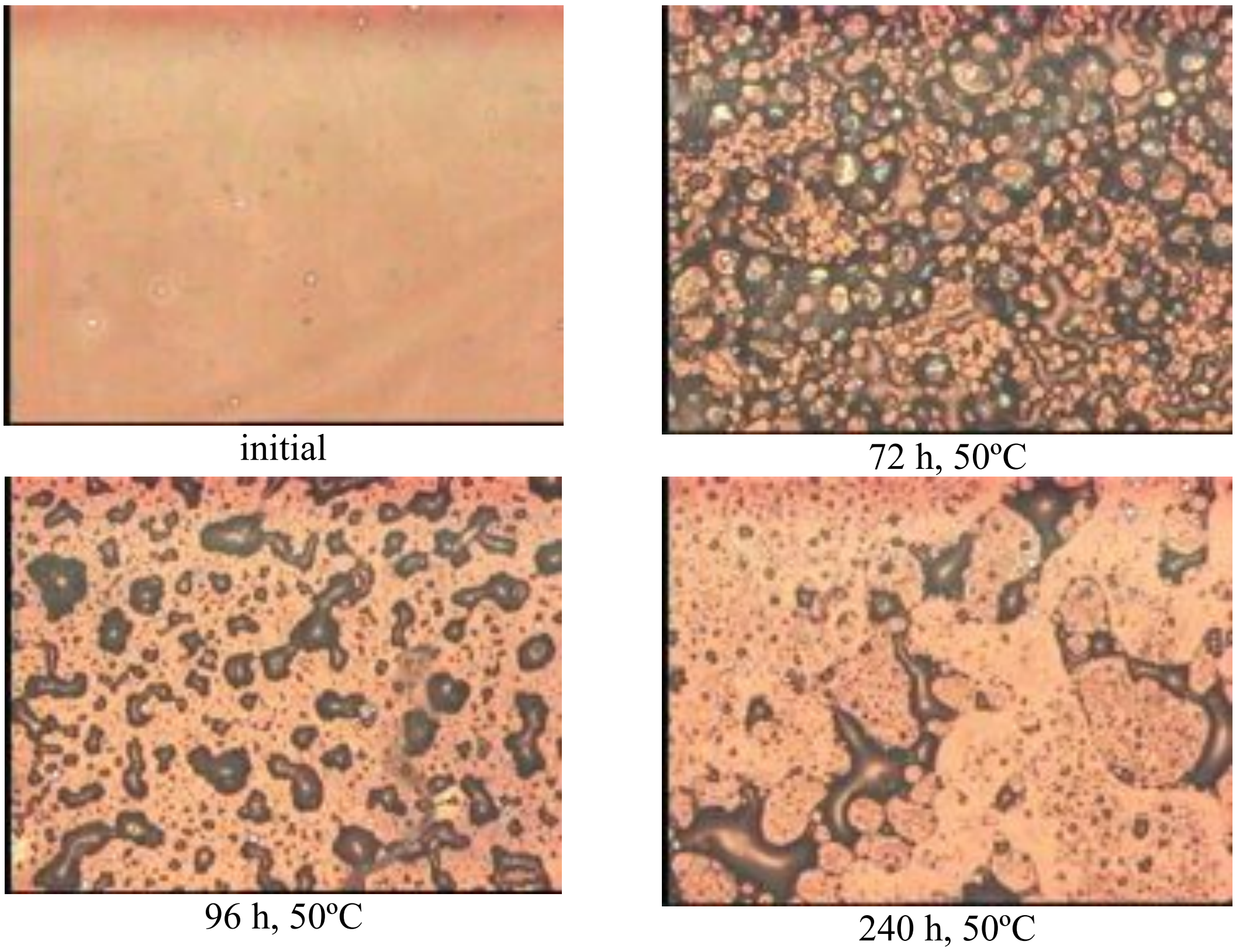

Fig.13. Microphotos of 1 μm untreated PLGA spincoated on carbon layer after water exposure at 50ºC. Size of photos is 1200x950 μm.

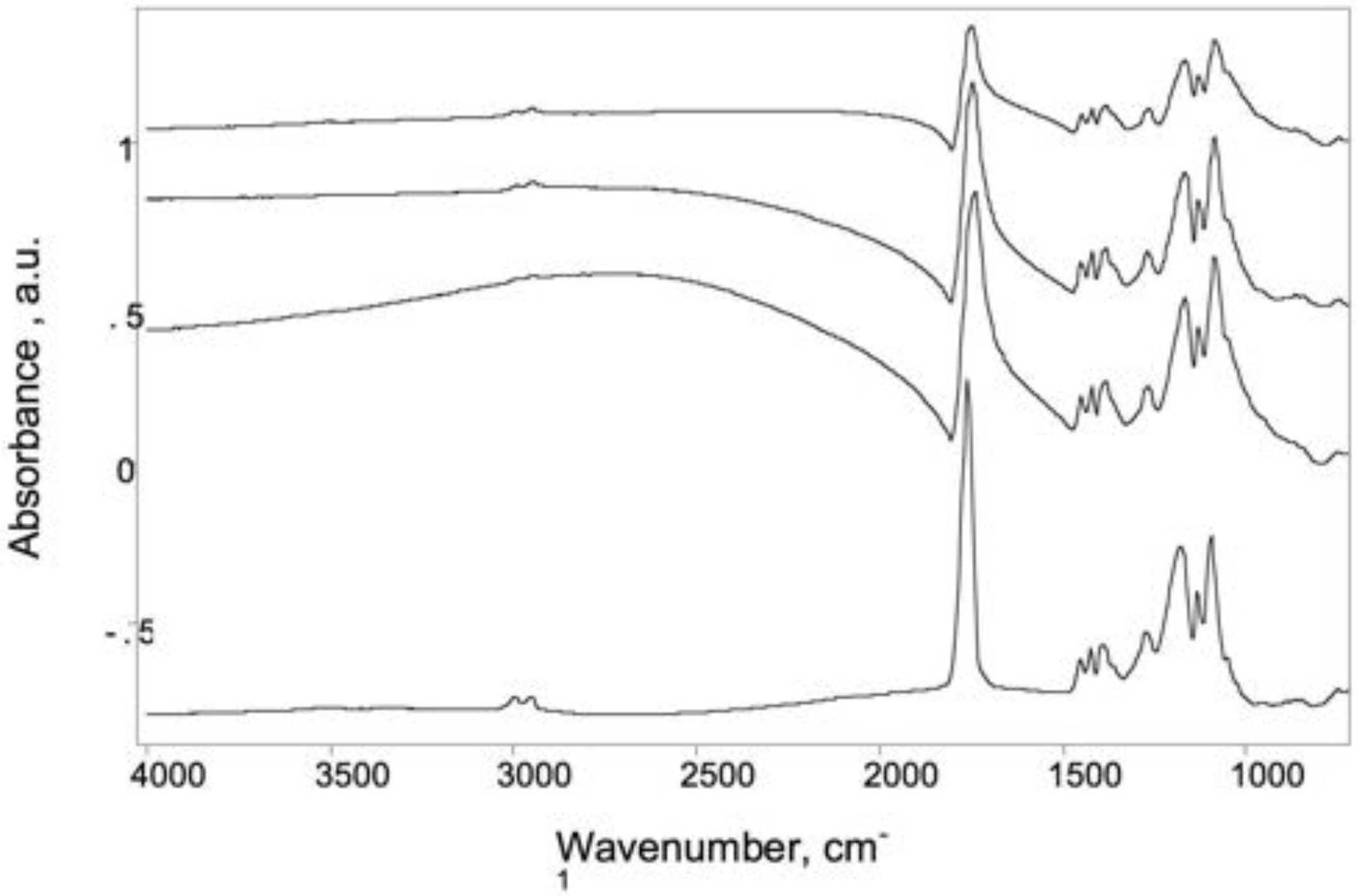

Fig.14. FTIR spectra of 1 mkm PLGA on carbon layer. From bottom: freshly spun, 24h, 48h, 72h in water at 50ºC.

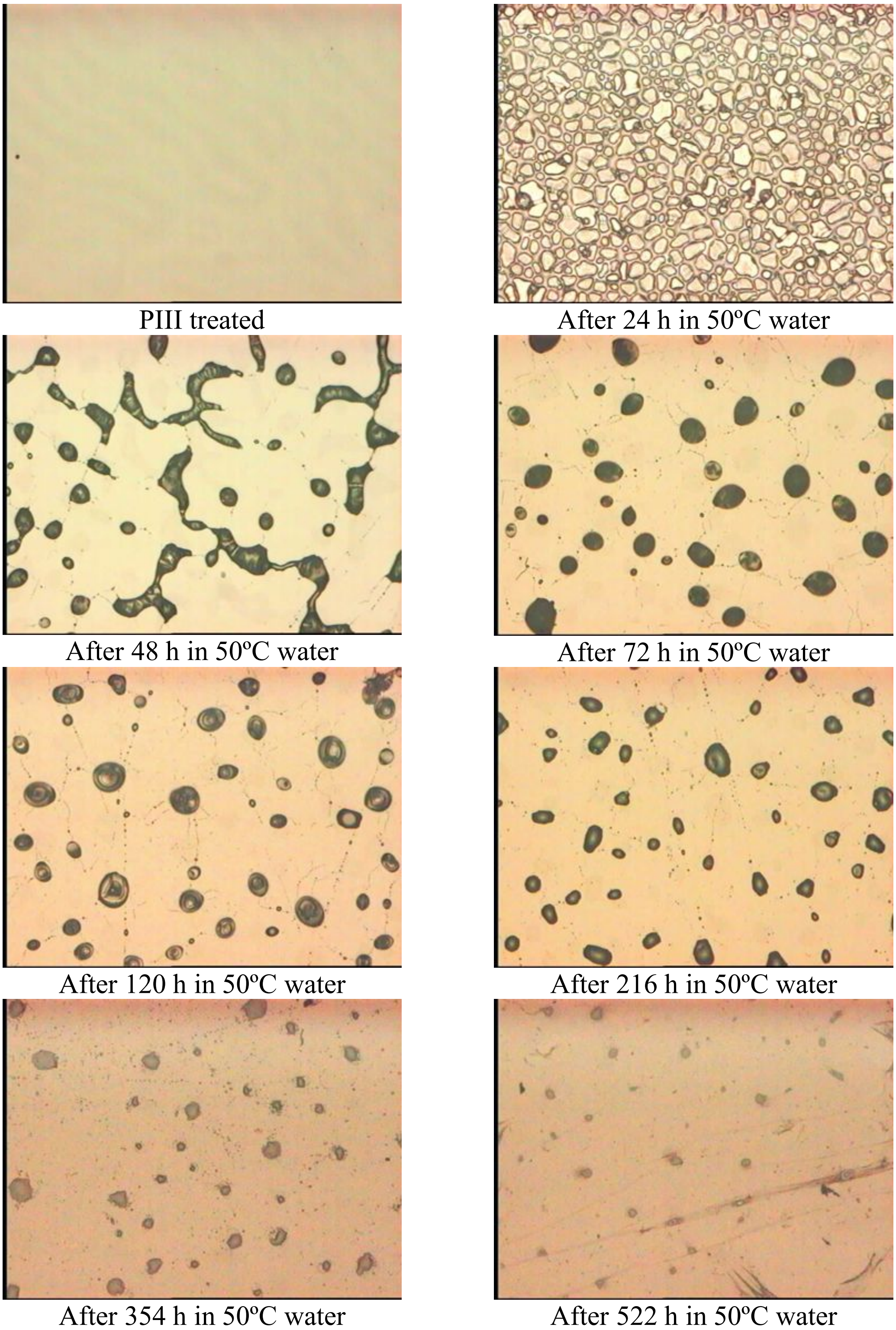

Fig.15. Microphotos of PLGA on silicon wafer after PIII at a fluence of $2\times10^{14}$ ions/cm$^2$. Size of photos is 1200x950 μm.

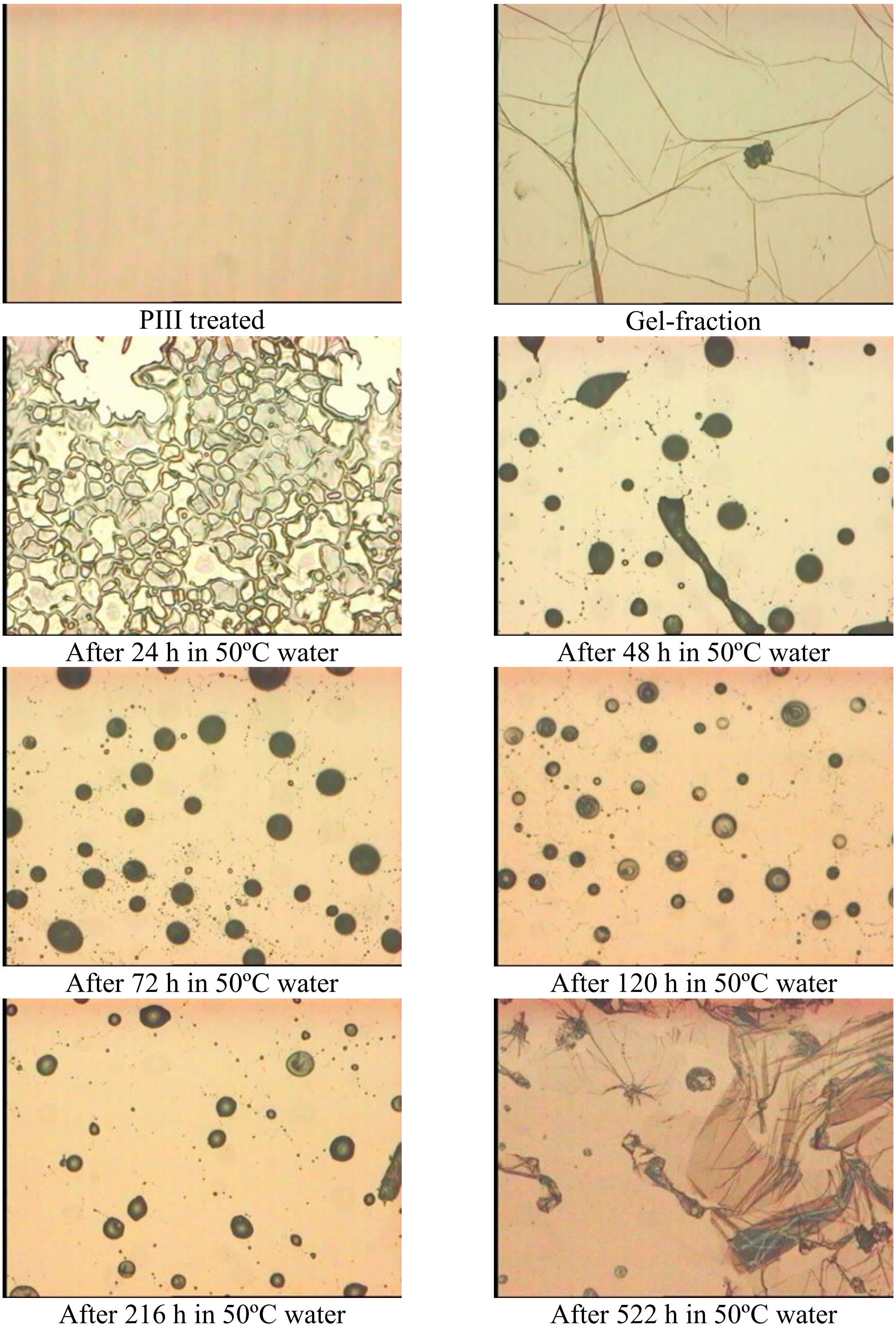

Fig.16. Microphotos of PLGA on silicon wafer treated with PIII to a fluence of $1\times10^{15}$ ions/cm$^2$. Size of photos is 1200x950 μm.

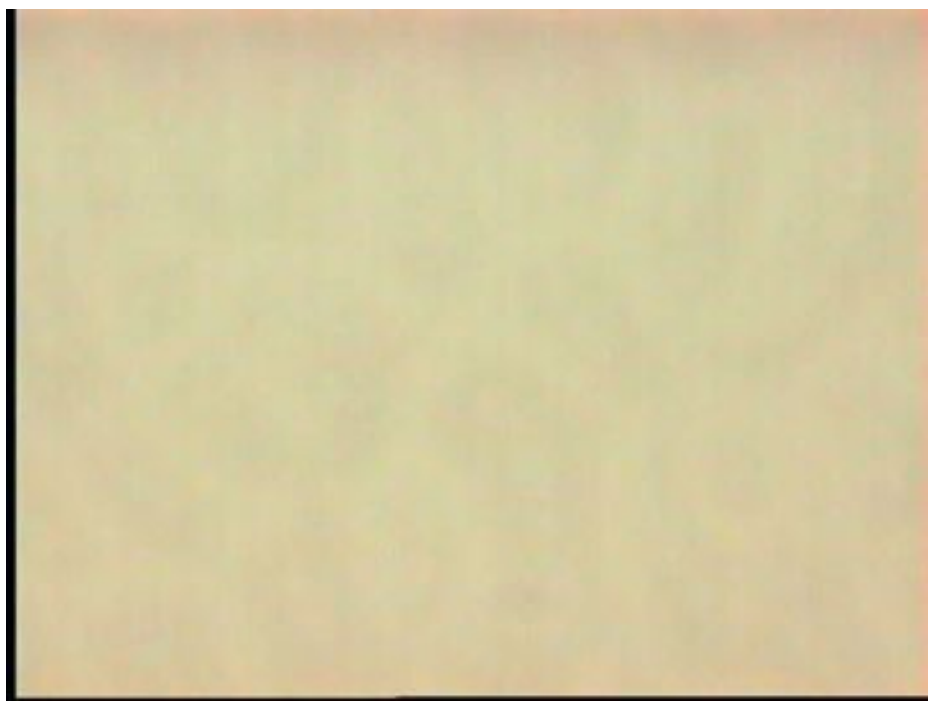

PIII treated

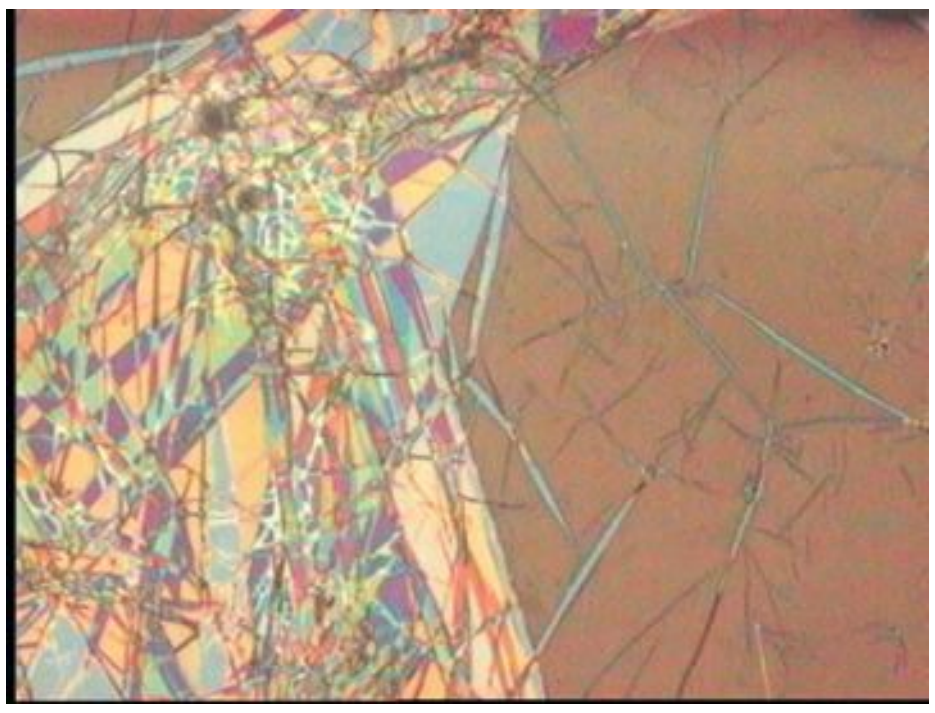

Gel-fraction

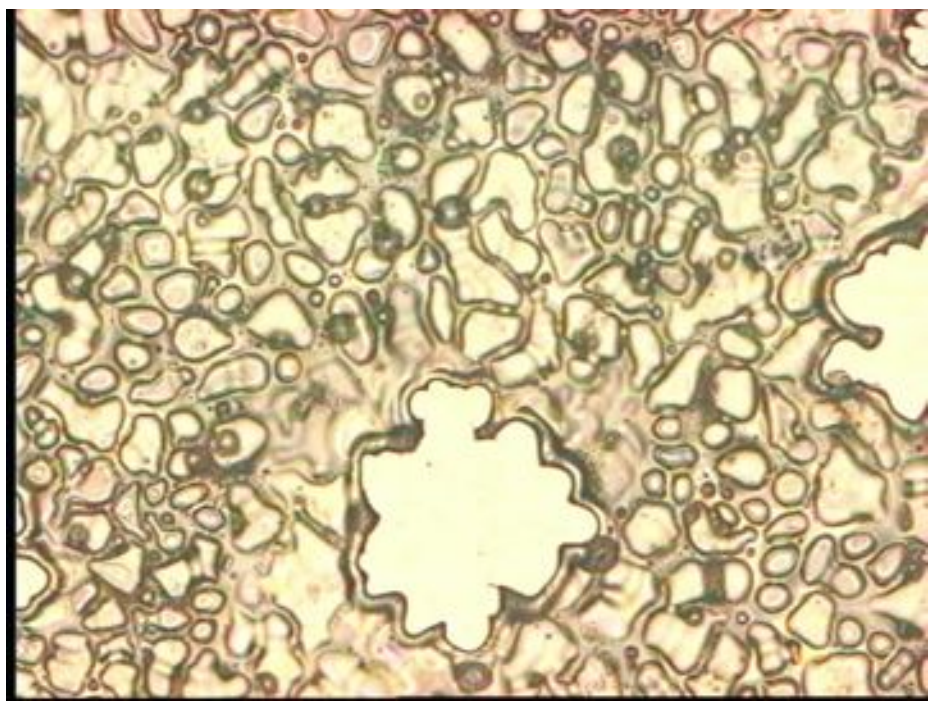

After 24 h in 50ºC water

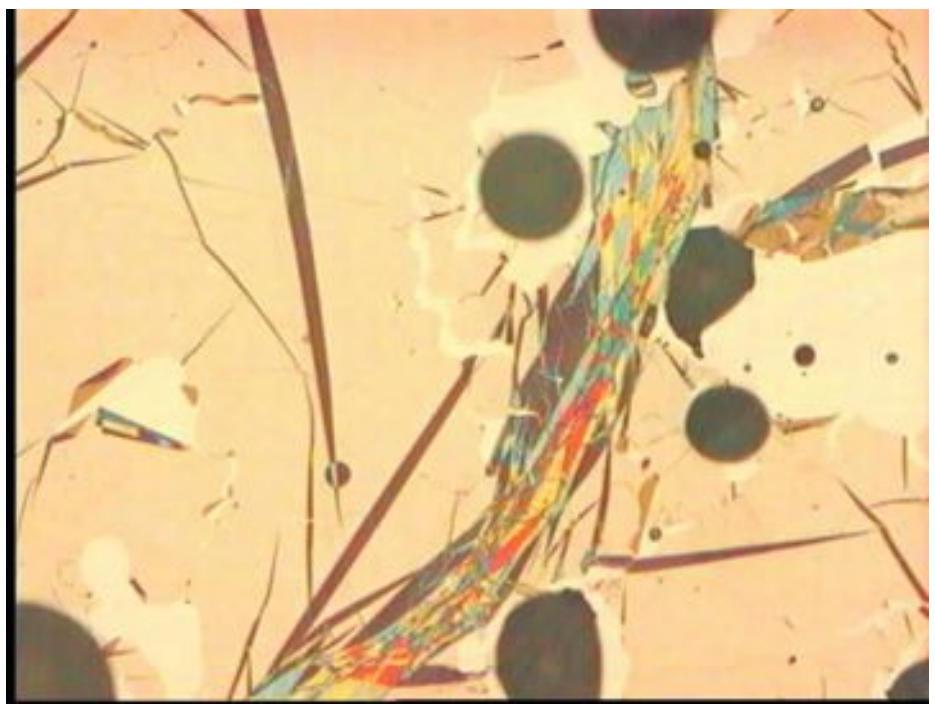

After 48 h in 50ºC water

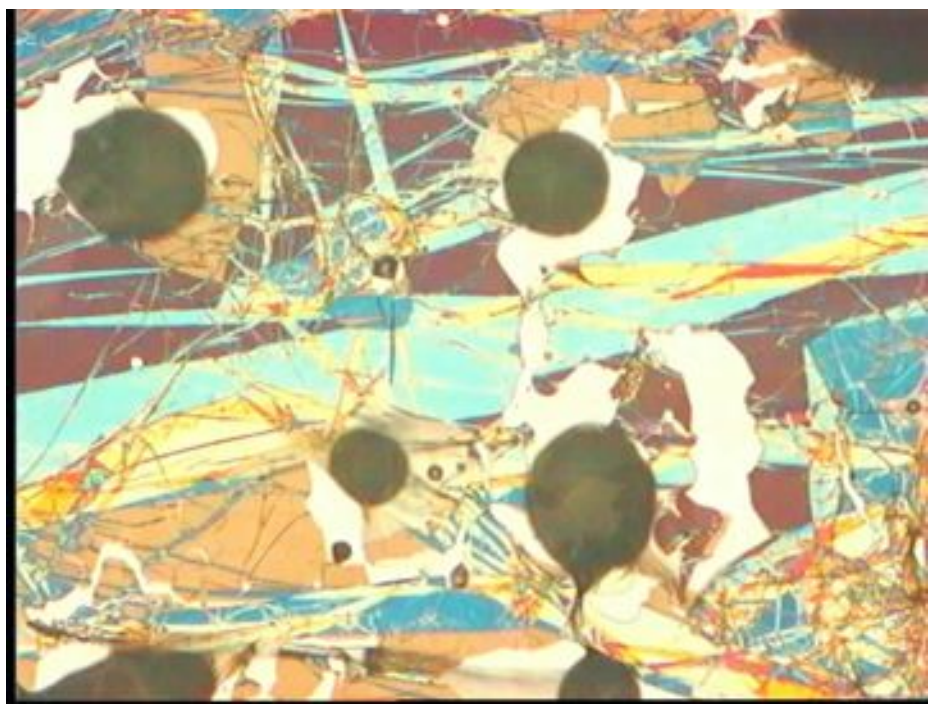

After 72 h in 50ºC water

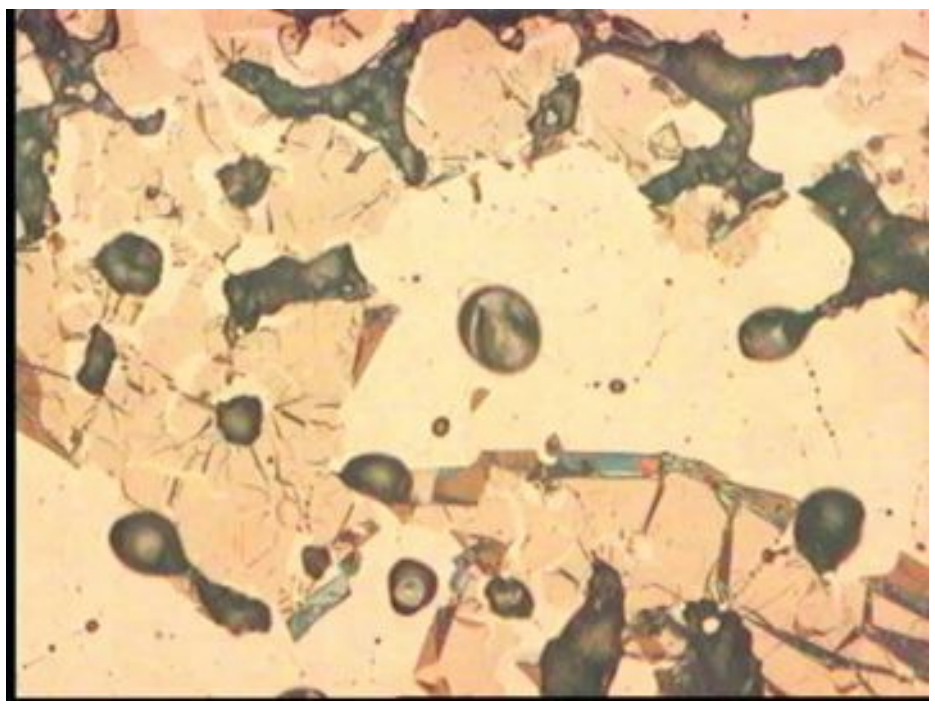

After 120 h in 50ºC water

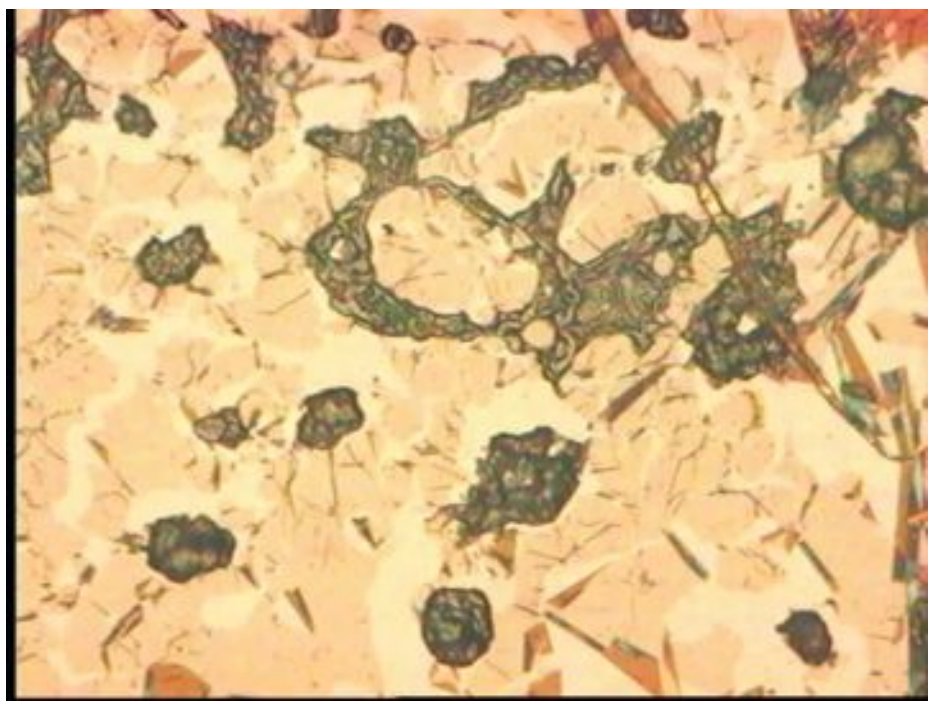

After 216 h in 50ºC water

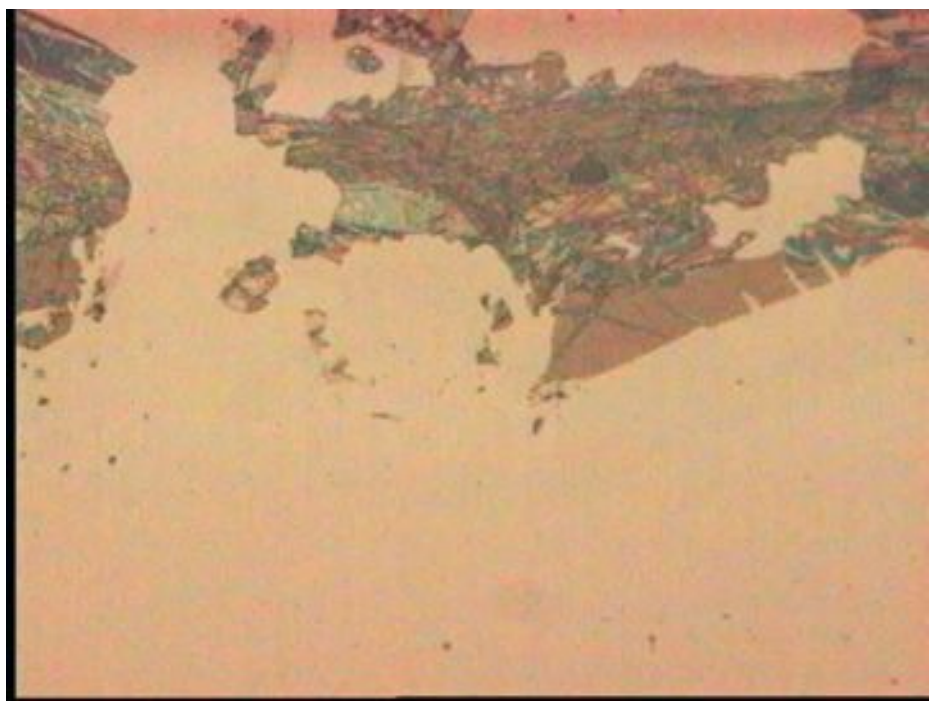

After 522 h in 50ºC water

Fig.17. Microphotos of PLGA on silicon after PIII with fluence $2\times10^{15}$ ions/cm$^2$. Size of photos is 1200x950 μm.

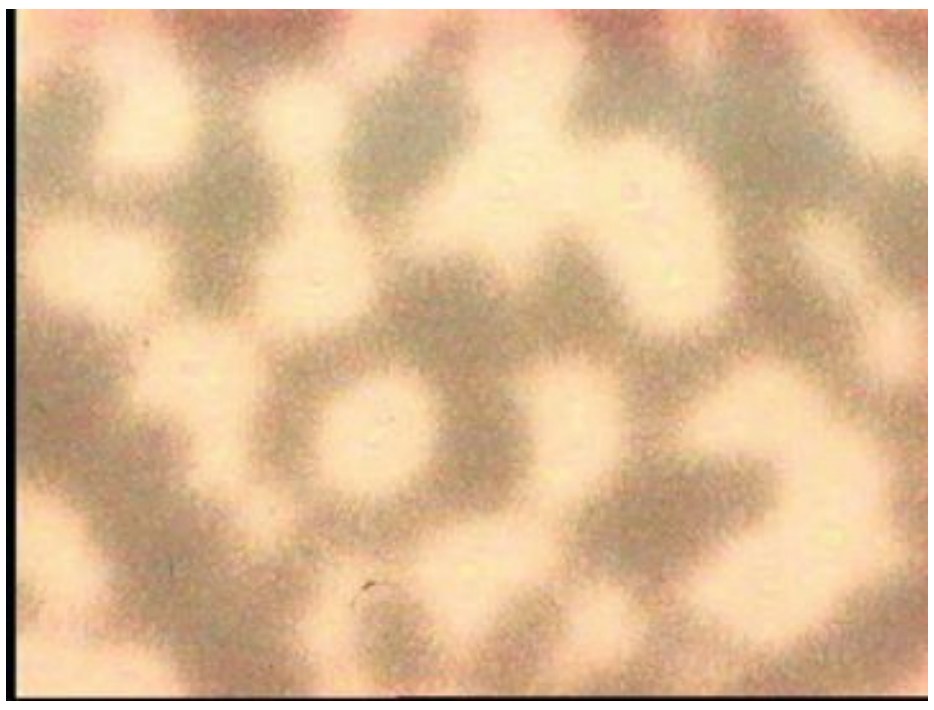

PIII treated

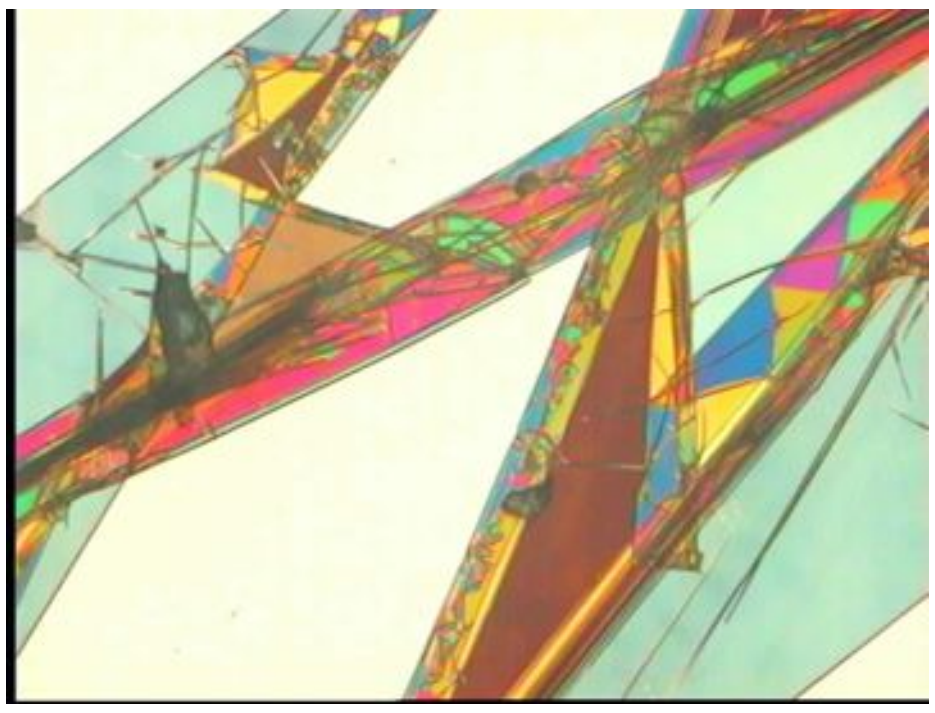

Gel-fraction

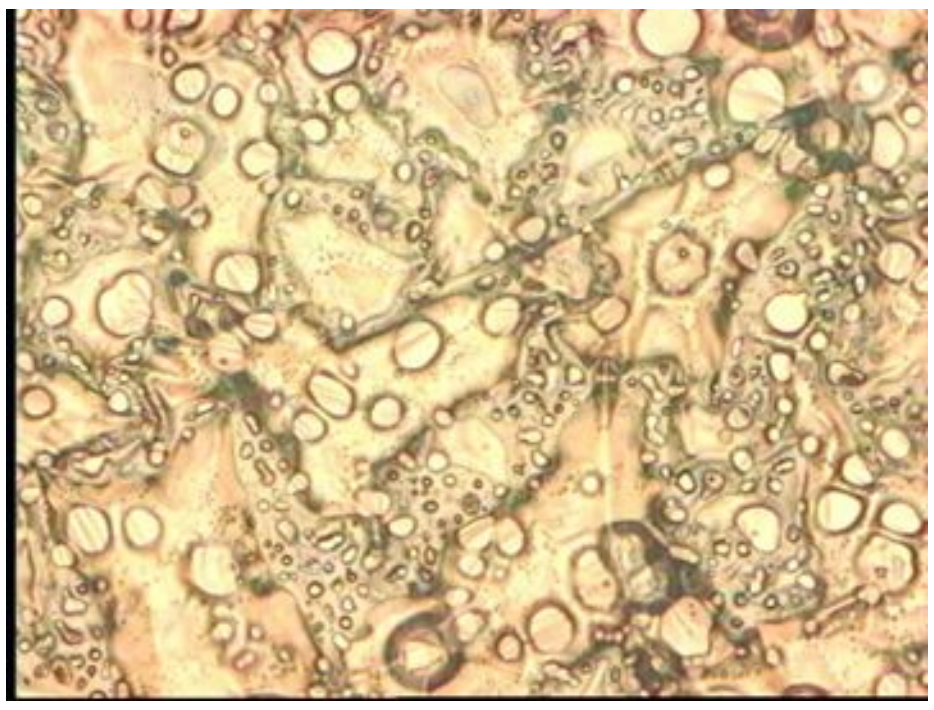

After 24 h in 50ºC water

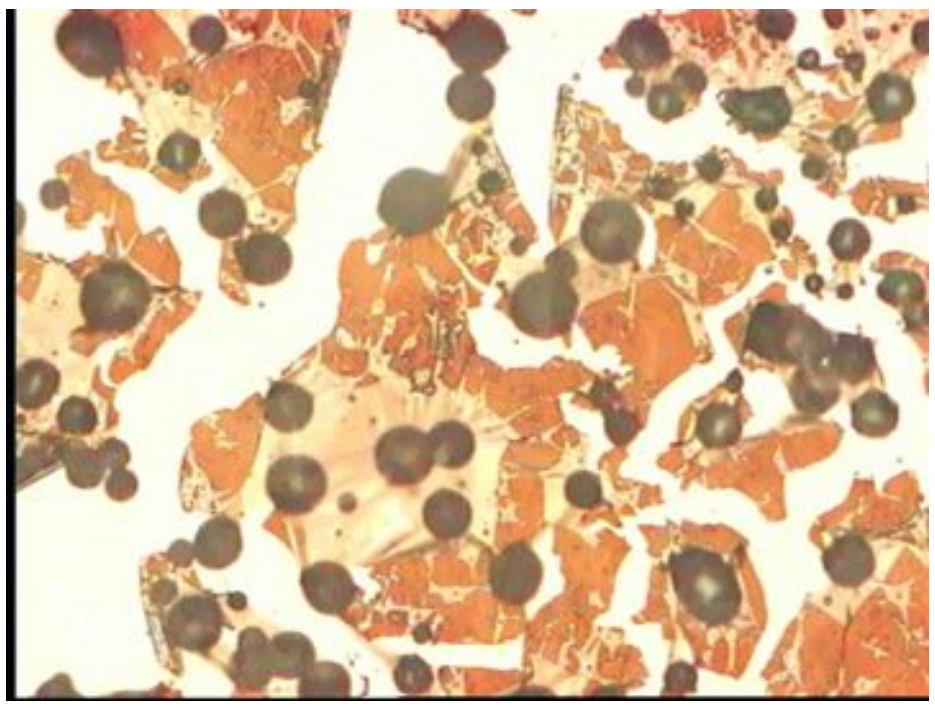

After 48 h in 50ºC water

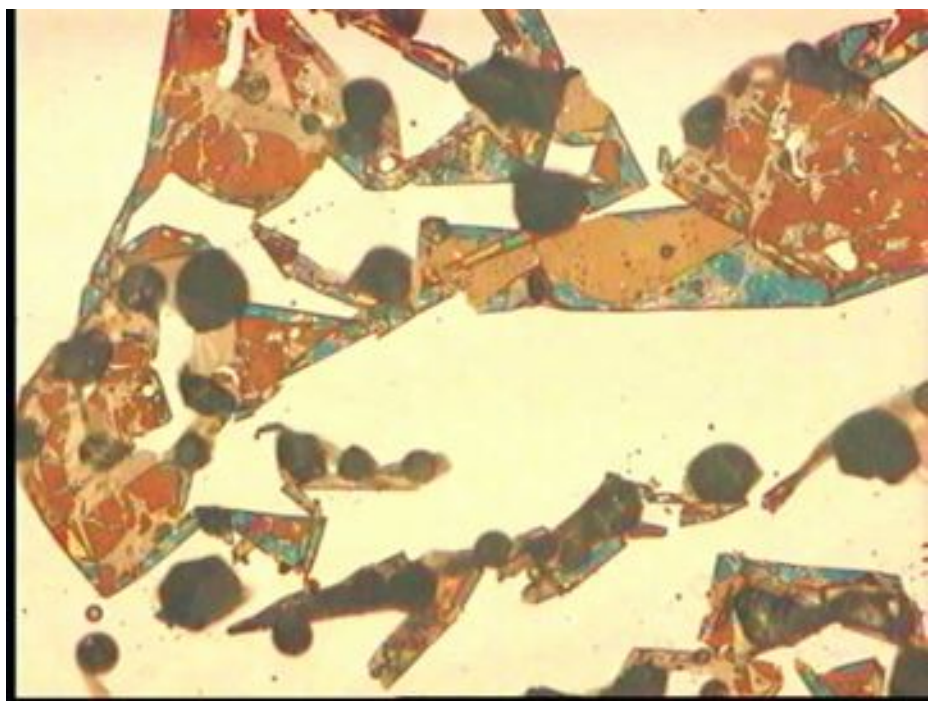

After 72 h in 50ºC water

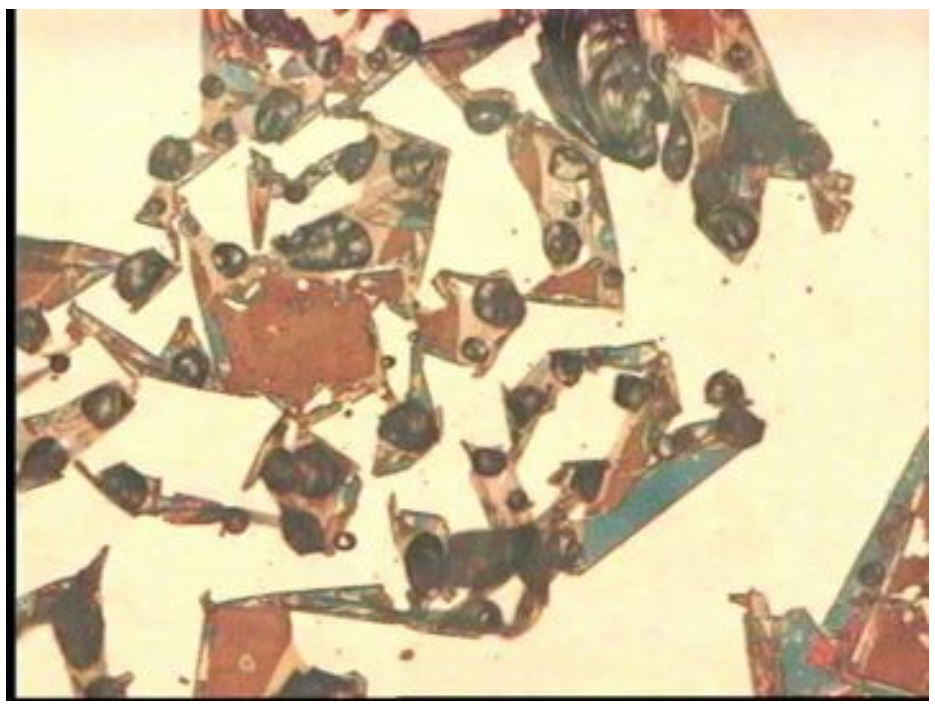

After 120 h in 50ºC water

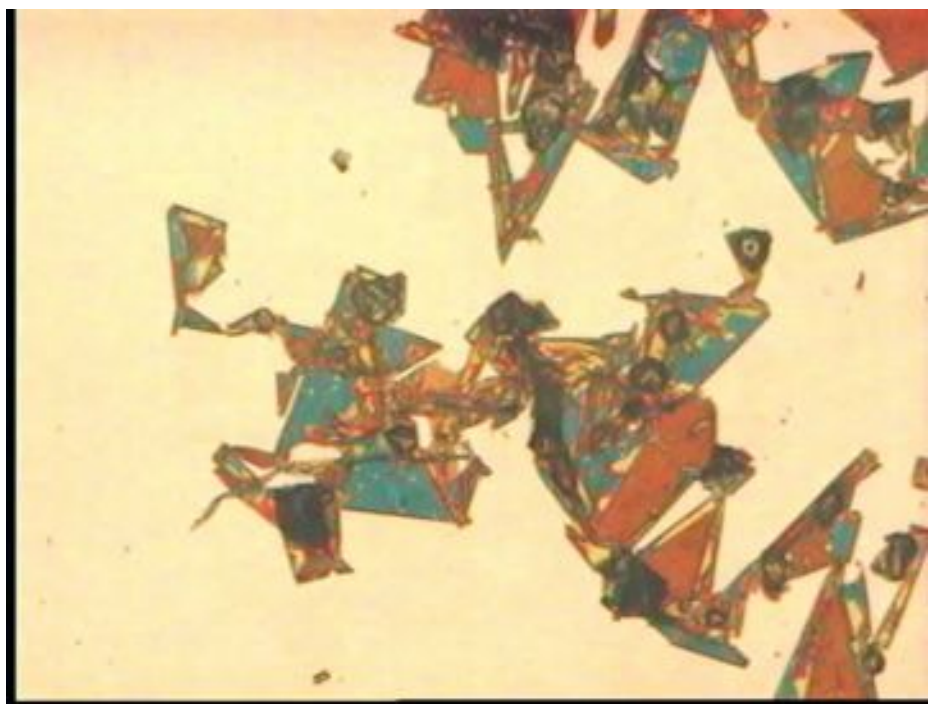

After 216 h in 50ºC water

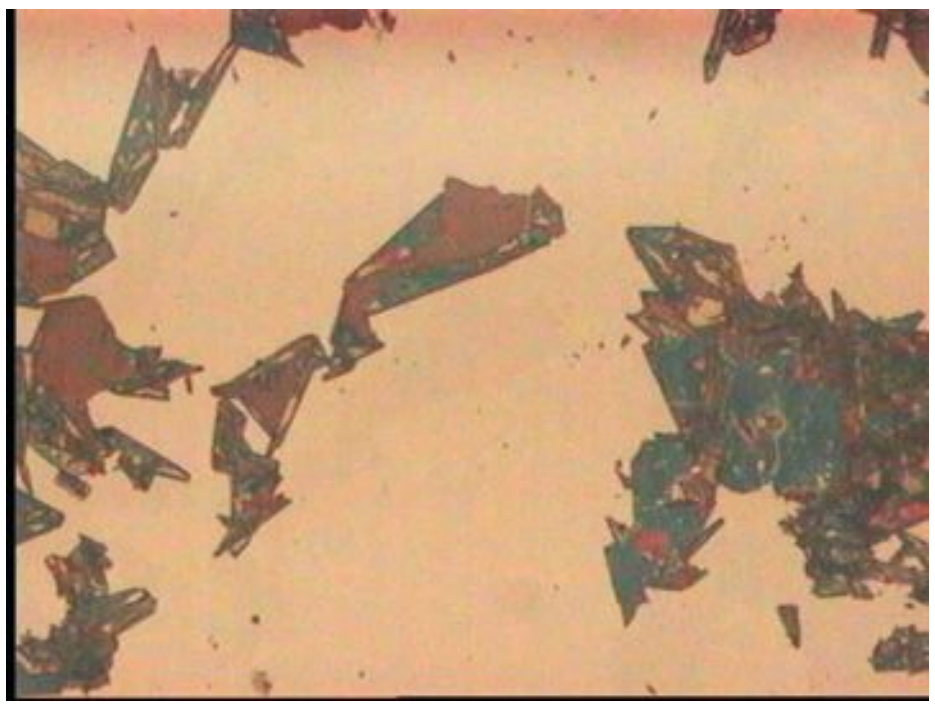

After 522 h in 50ºC water

Fig.18. Microphotos of PLGA on silicon after PIII treatment to a fluence of $1\times10^{16}$ ions/$cm^2$. Size of photos is 1200x950 μm.

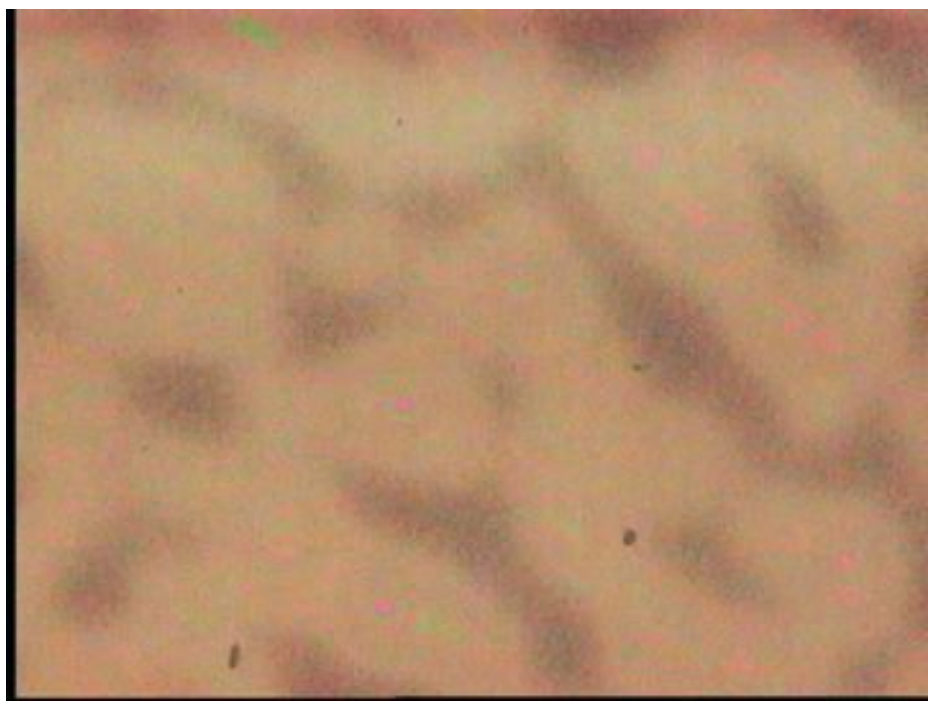

PIII treated

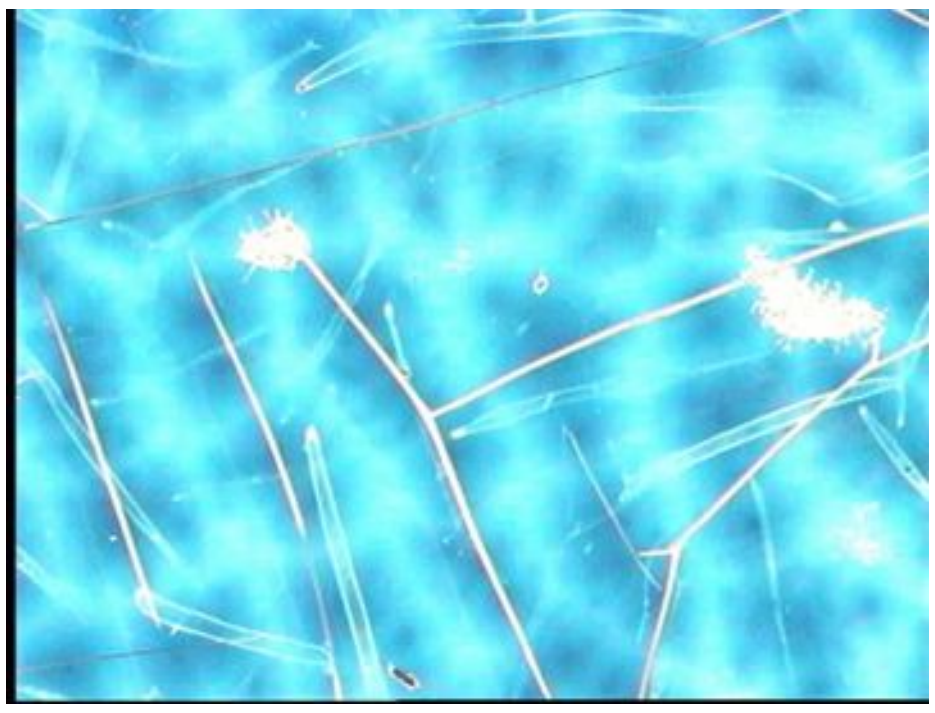

Gel-fraction

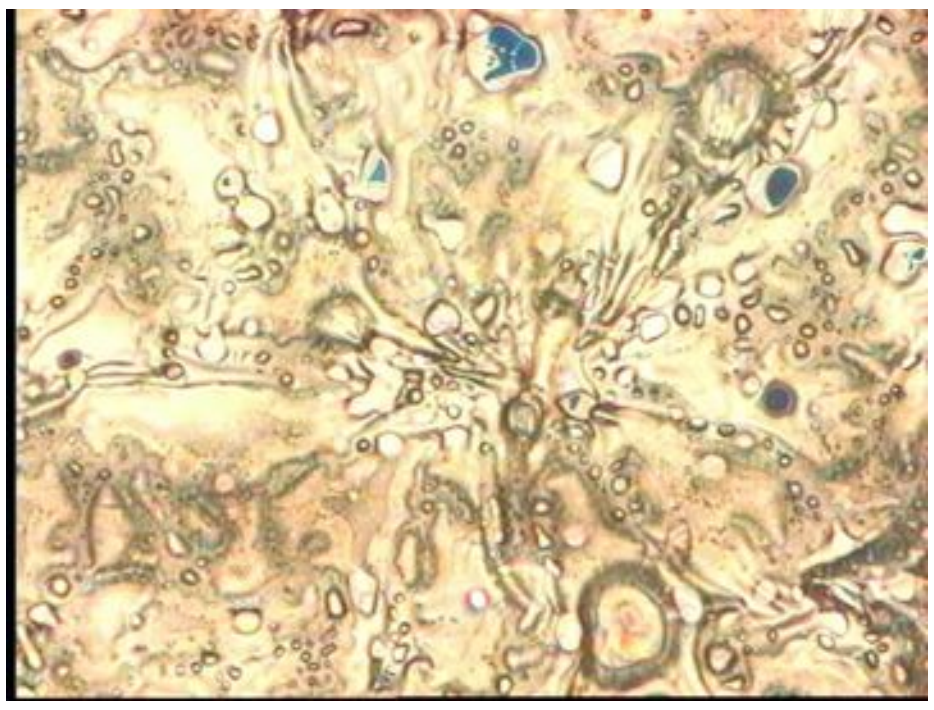

After 24 h in 50ºC water

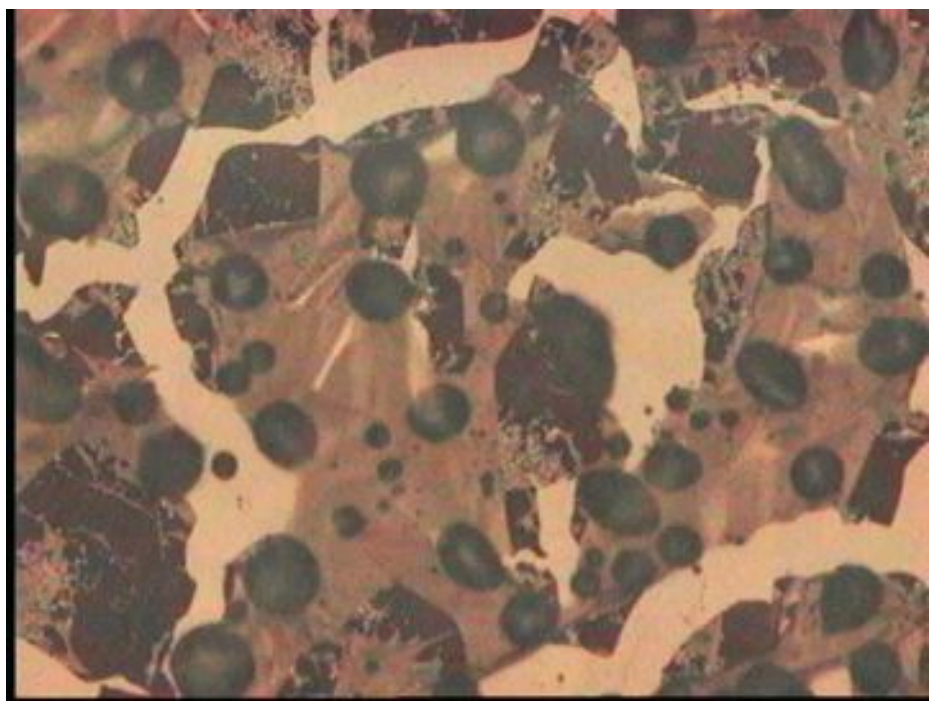

After 48 h in 50ºC water

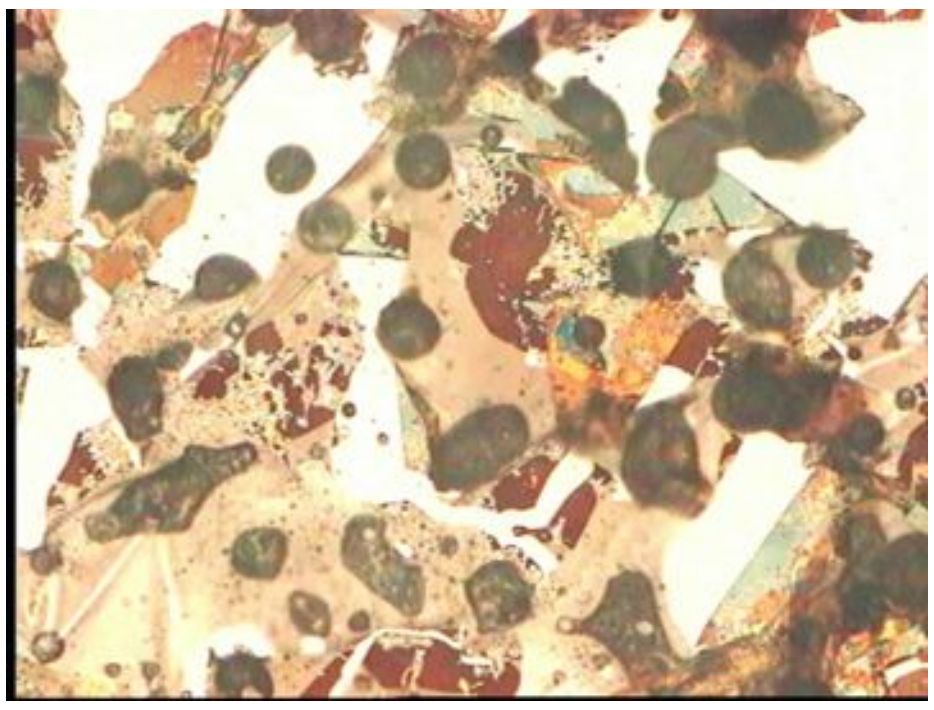

After 72 h in 50ºC water

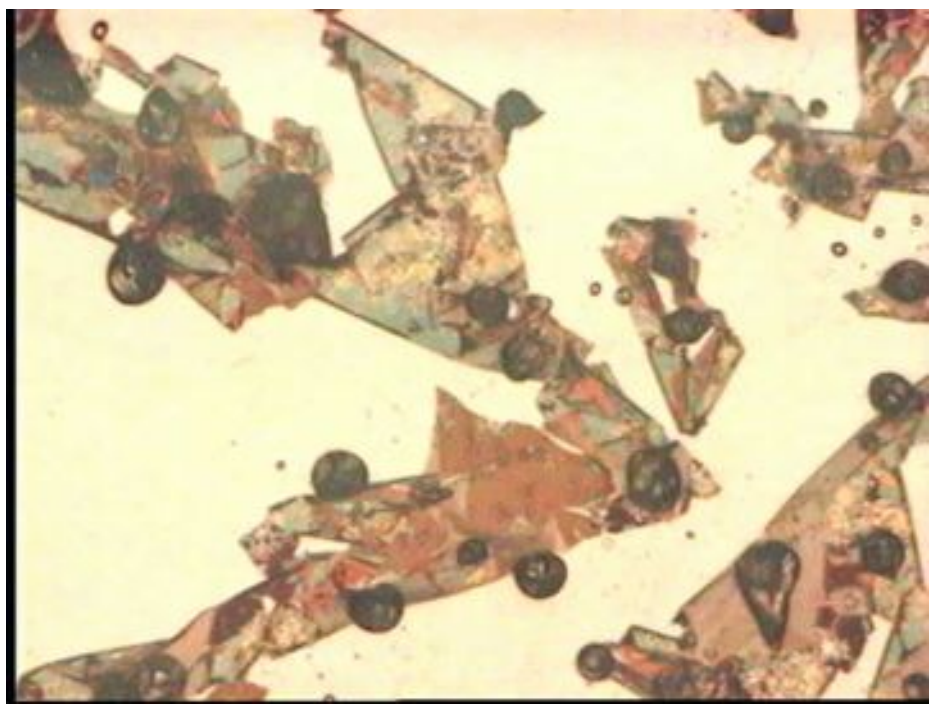

After 120 h in 50ºC water

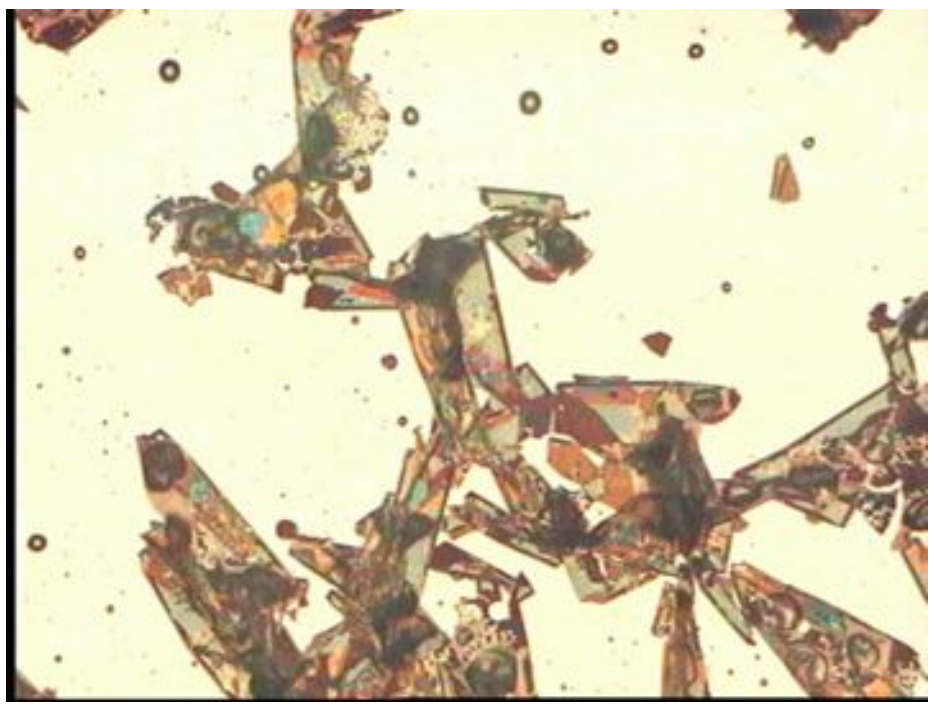

After 216 h in 50ºC water

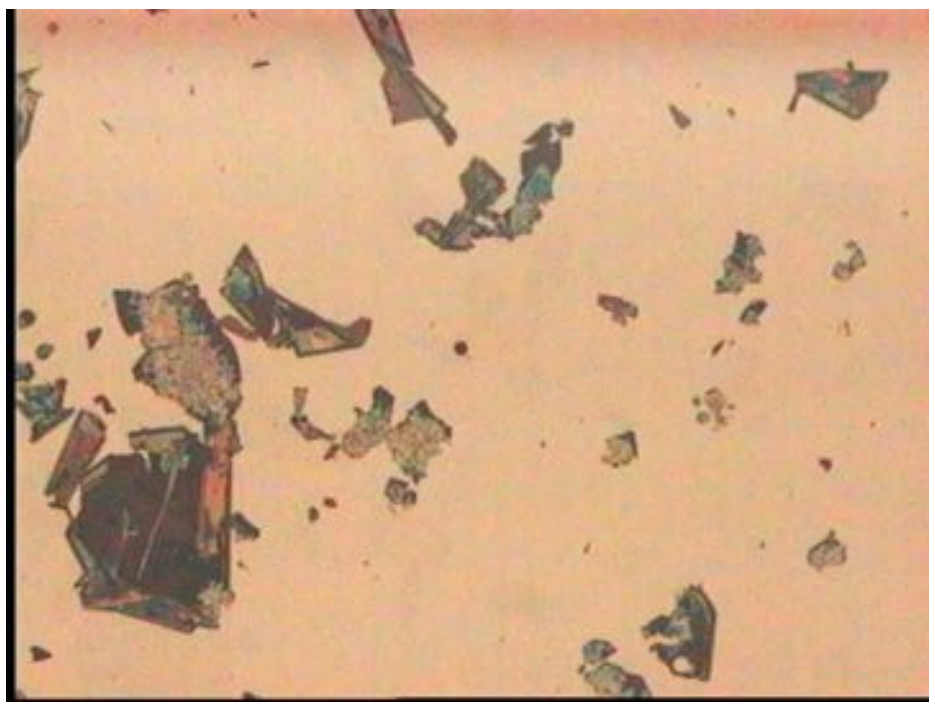

After 522 h in 50ºC water

Fig.19. Microphotos of PLGA on silicon after PIII treatment to a fluence of $2\times10^{16}$ ions/cm$^2$. Size of photos is 1200x950 μm.

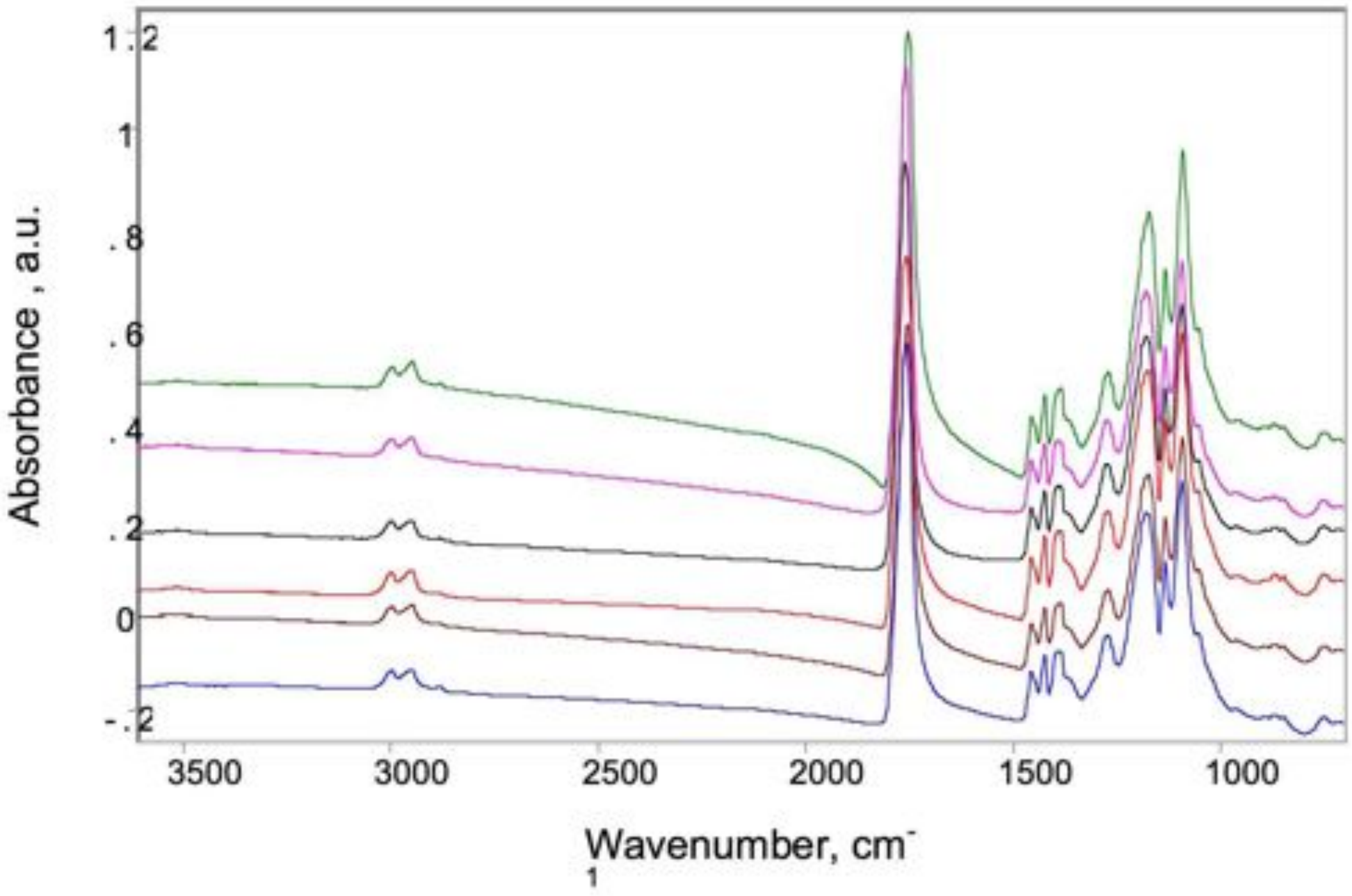

Fig.20. FTIR spectra of PIII treated 1 μm PLGA after 24 h exposure to water at 50ºC. The curves correspond to samples treated with the following ion fluences: green – untreated, pink – 2×10$^{14}$, black – 10$^{15}$, red – 2×10$^{15}$, brown – 10$^{16}$, blue – 2×10$^{16}$ ions/cm$^2$.

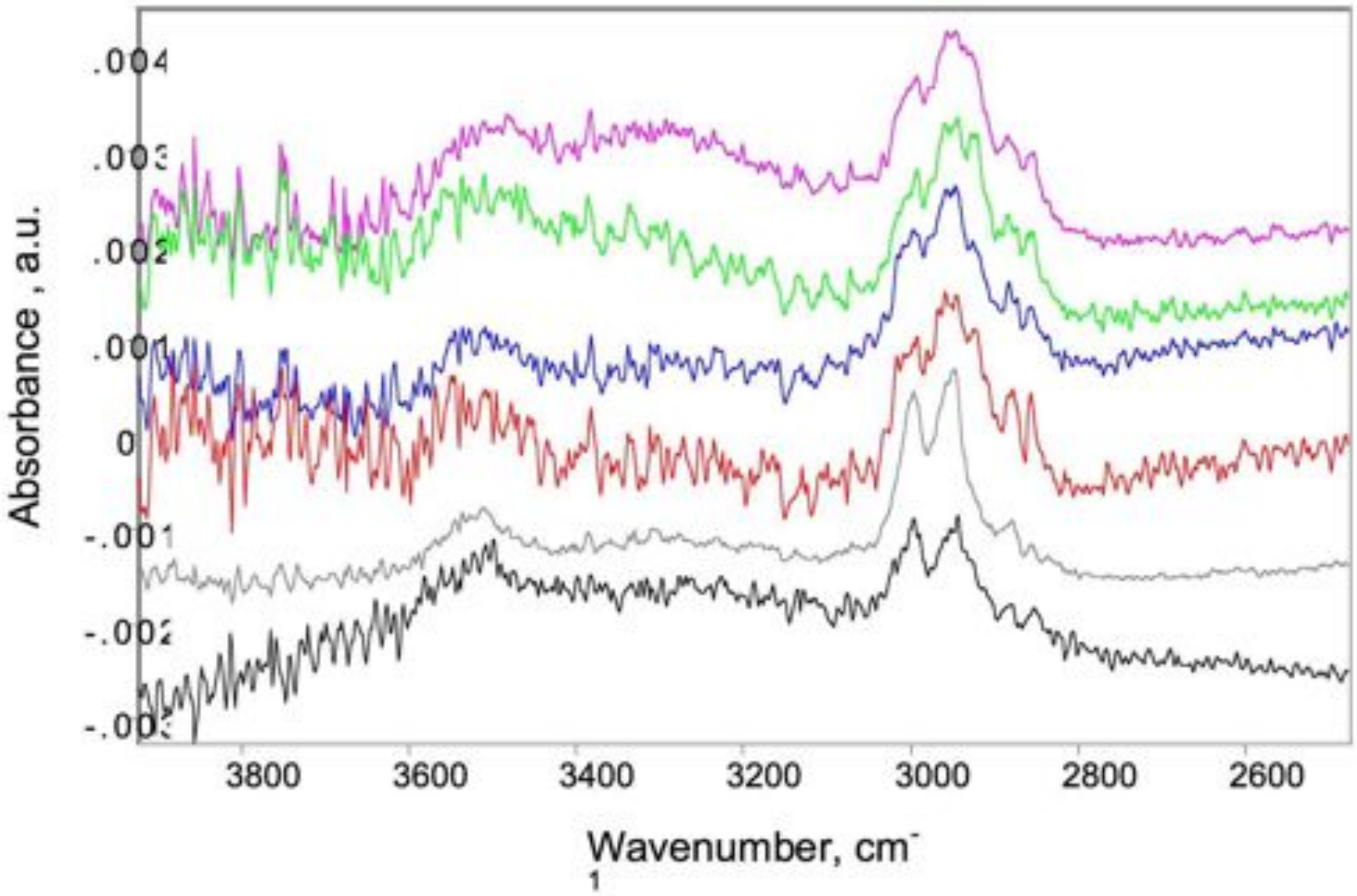

Fig.21. FTIR spectra of 1 μm PIII treated PLGA exposed for 72 hours in water at 50ºC. The curves show the spectra for samples treated with the following fluences: black – untreated, gray – 2×10$^{14}$, red – 10$^{15}$, blue – 2×10$^{15}$, green – 10$^{16}$, pink – 2×10$^{16}$ ions/cm$^2$.

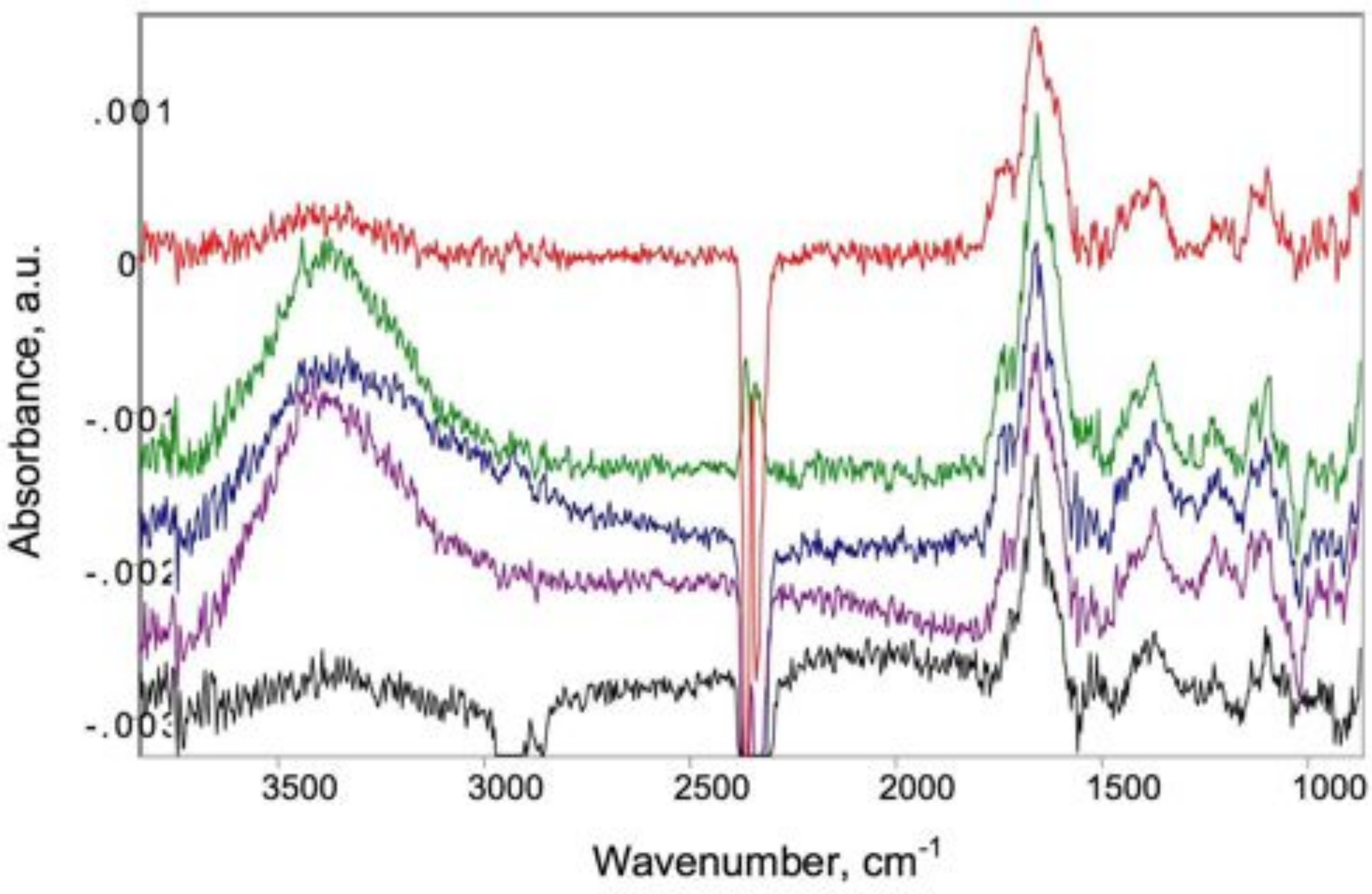

Fig.22. FTIR ATR spectra of degradation products from PLGA after PIII fluences of from bottom: black – untreated, violet – $10^{15}$, blue - $2\times10^{15}$, green – $10^{16}$, red – $2\times10^{16}$ ions/cm$^2$.

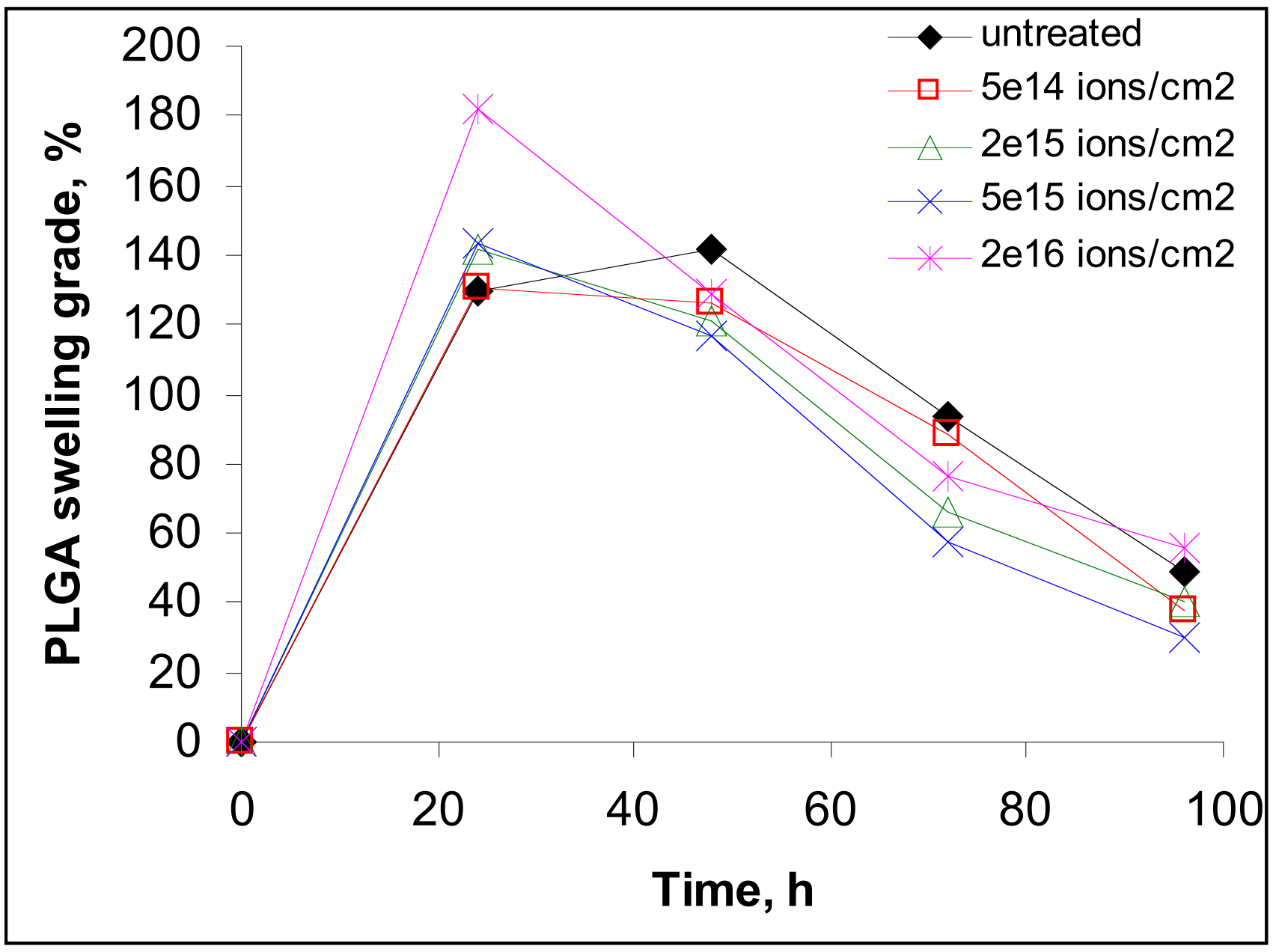

Fig.23. Swelling kinetics of 15 μm PLGA film in water at different PIII fluence.

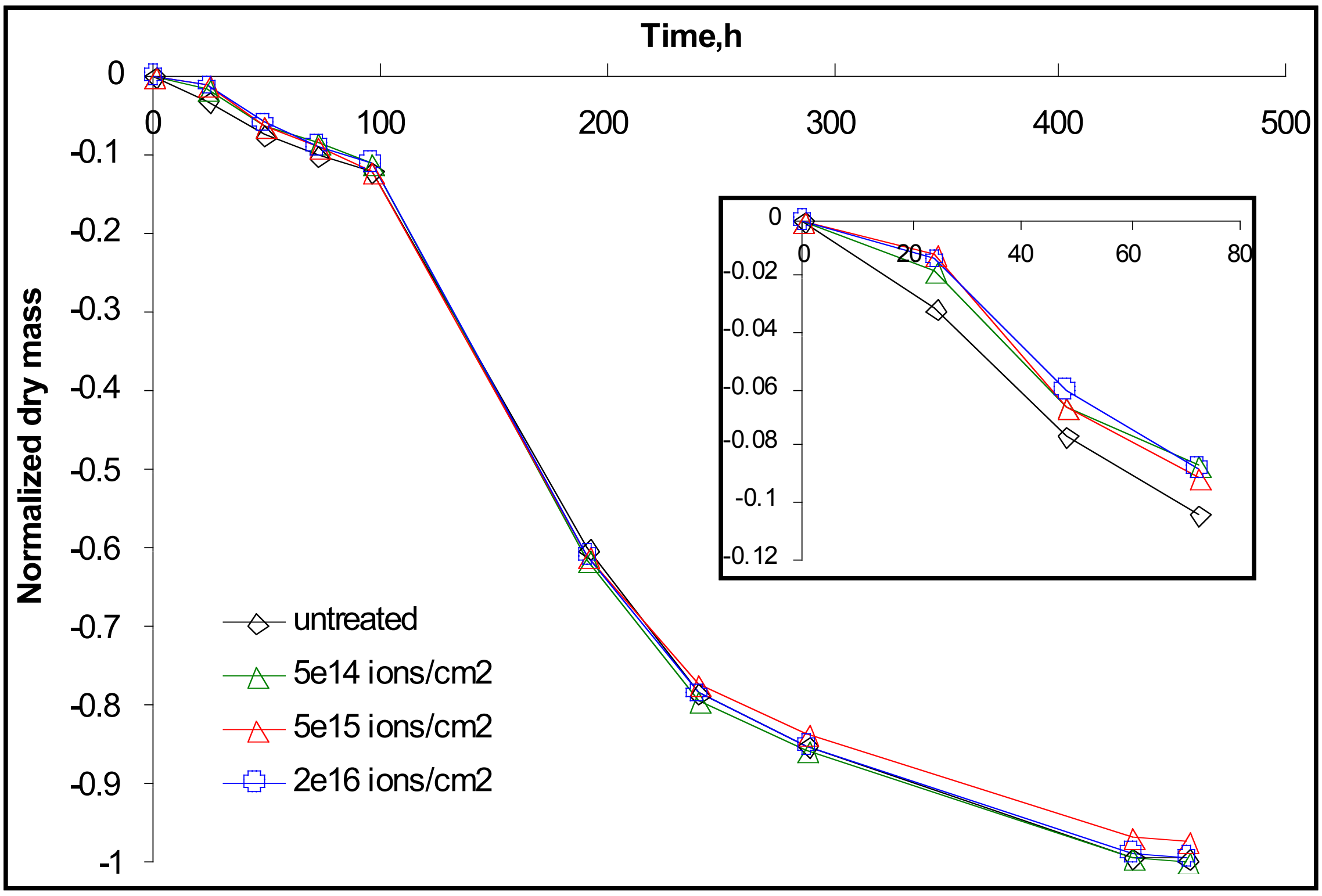

Fig.24. Mass loss of PLGA in water (50ºC). Fluence of PIII treatment is noted in box label.

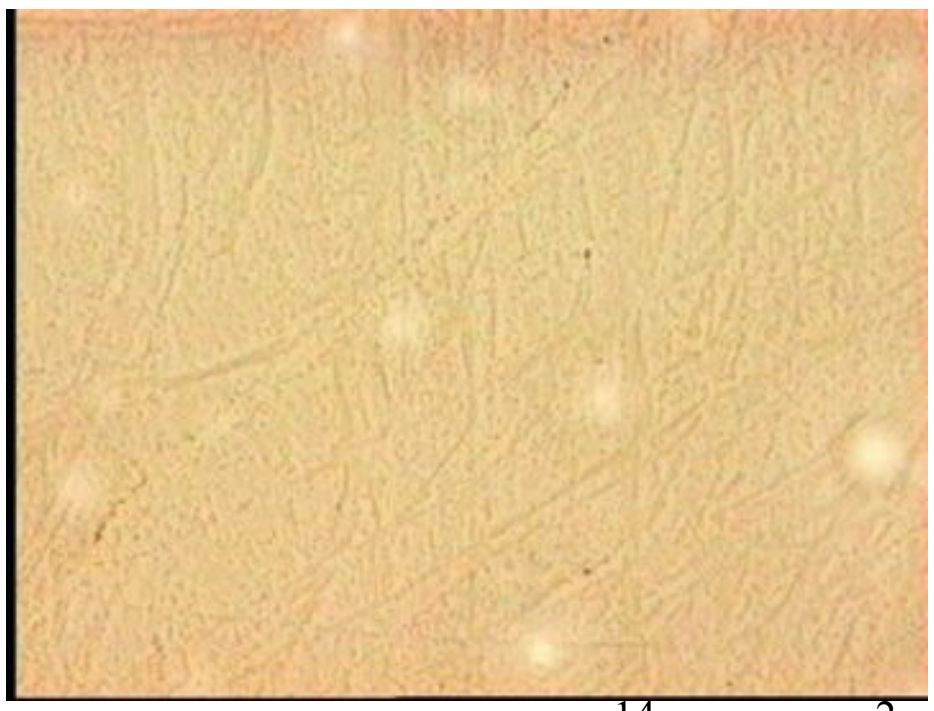

PIII treated, 20 kV, $10^{14}$ ions/cm$^2$

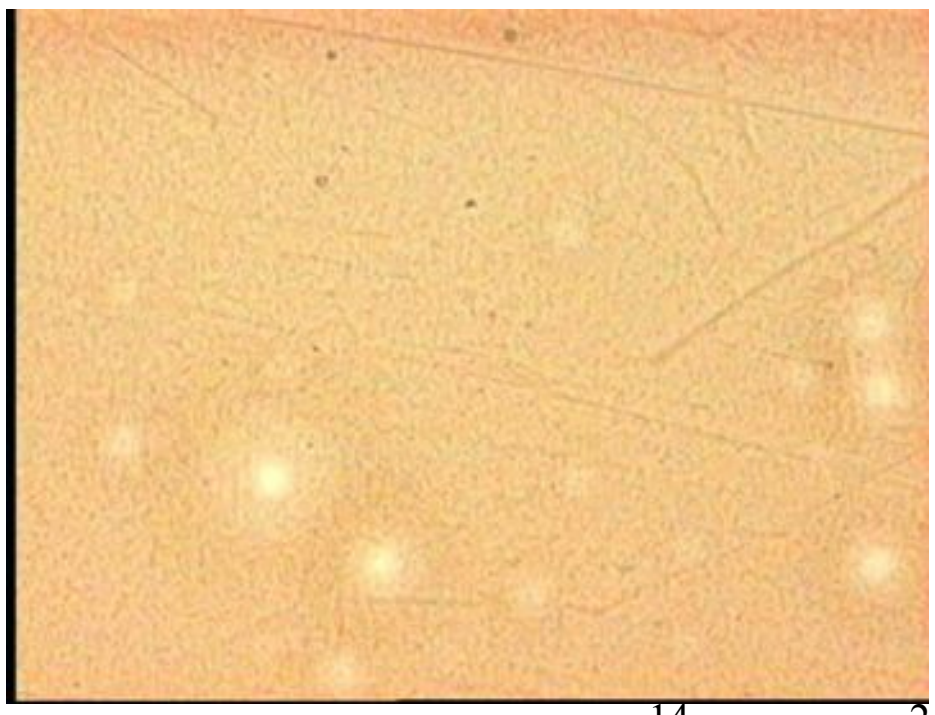

PIII treated, 10 kV, $5\times10^{14}$ ions/cm$^2$

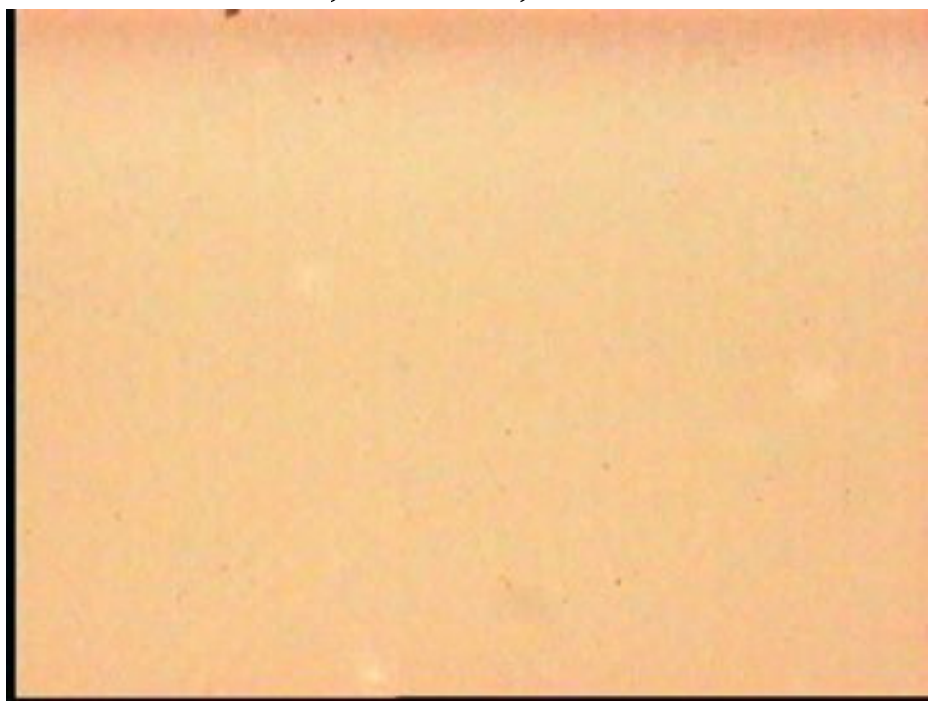

1 day incubation

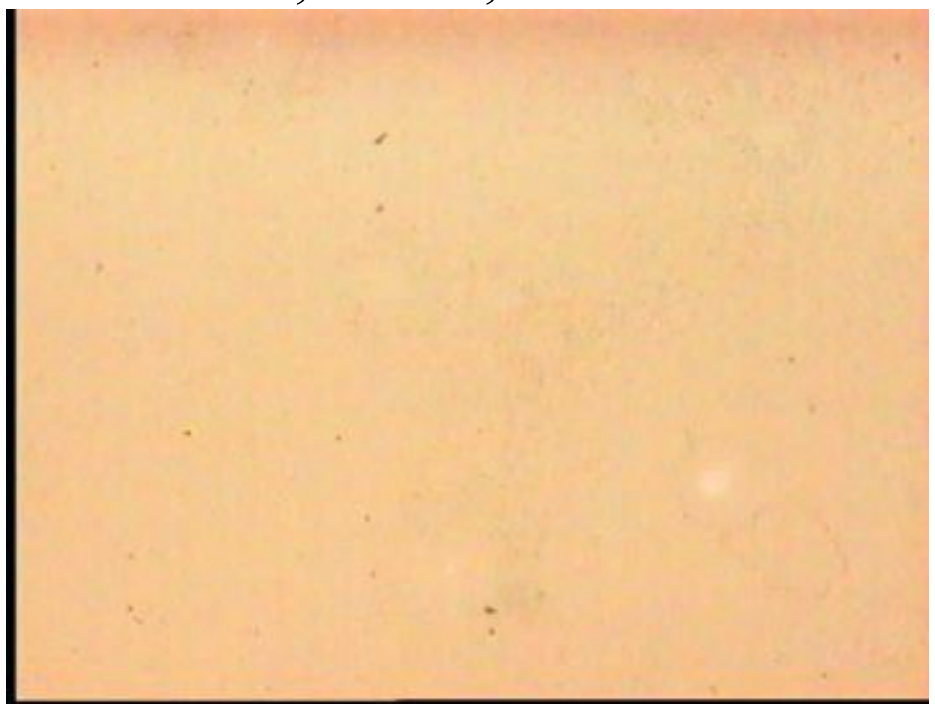

1 day incubation

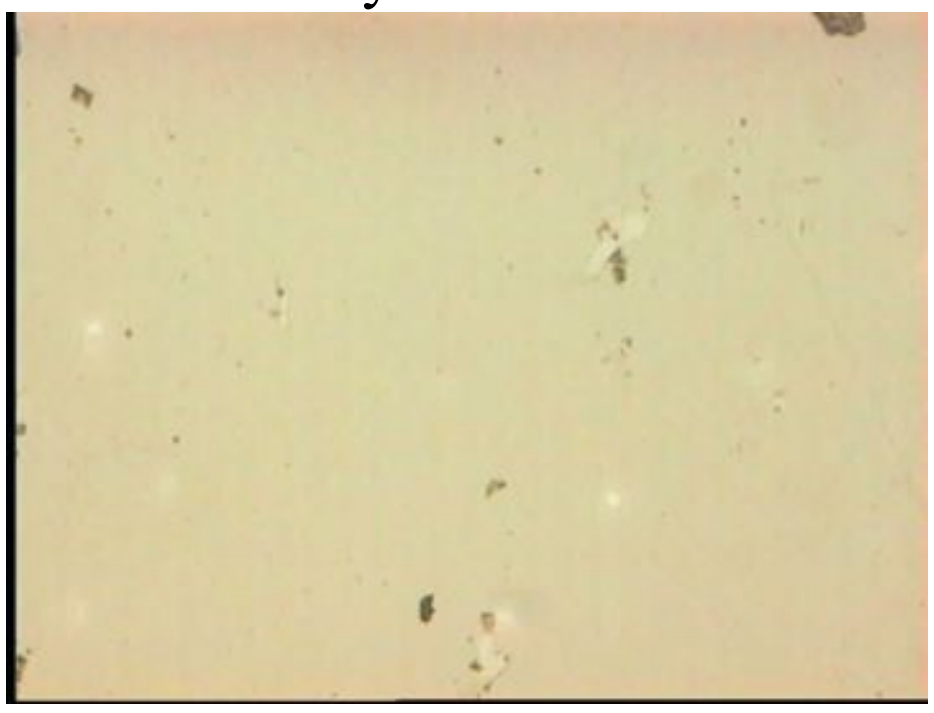

4 days incubation

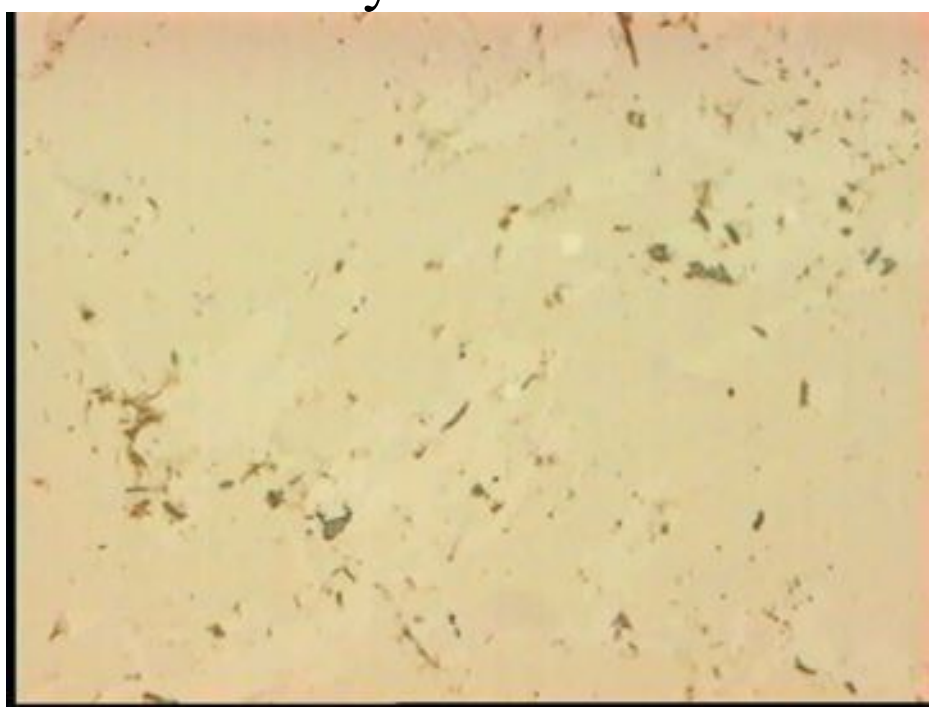

4 days incubation

Fig.25 Microphotos of 100 nm PLGA coating on silicon after 20 and 10 keV PIII treatment and incubation in water at 50ºC. Size of photos is 1200x950 μm.

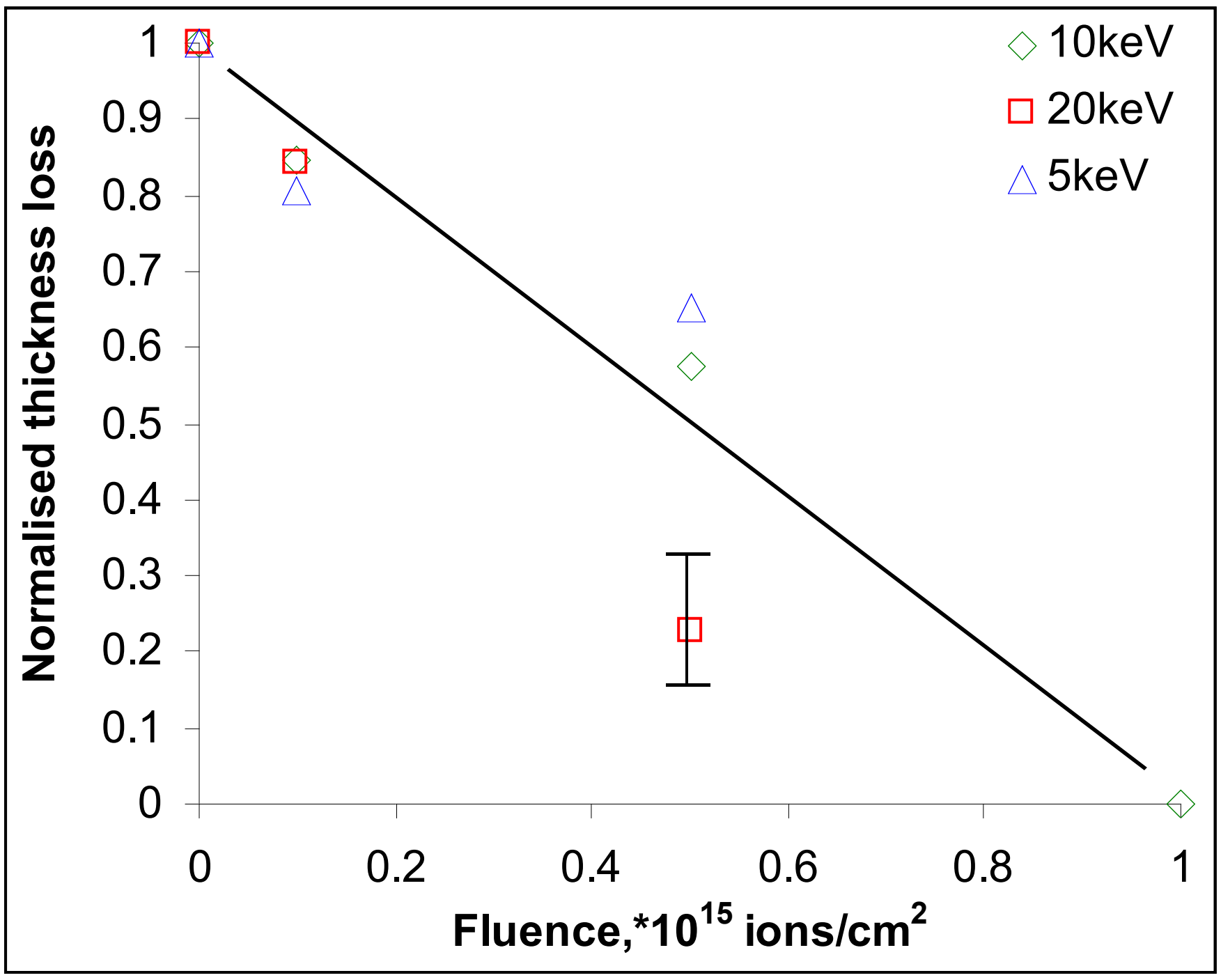

Fig.26. Degradable part of 100 nm PLGA coating in water after PIII treatment.

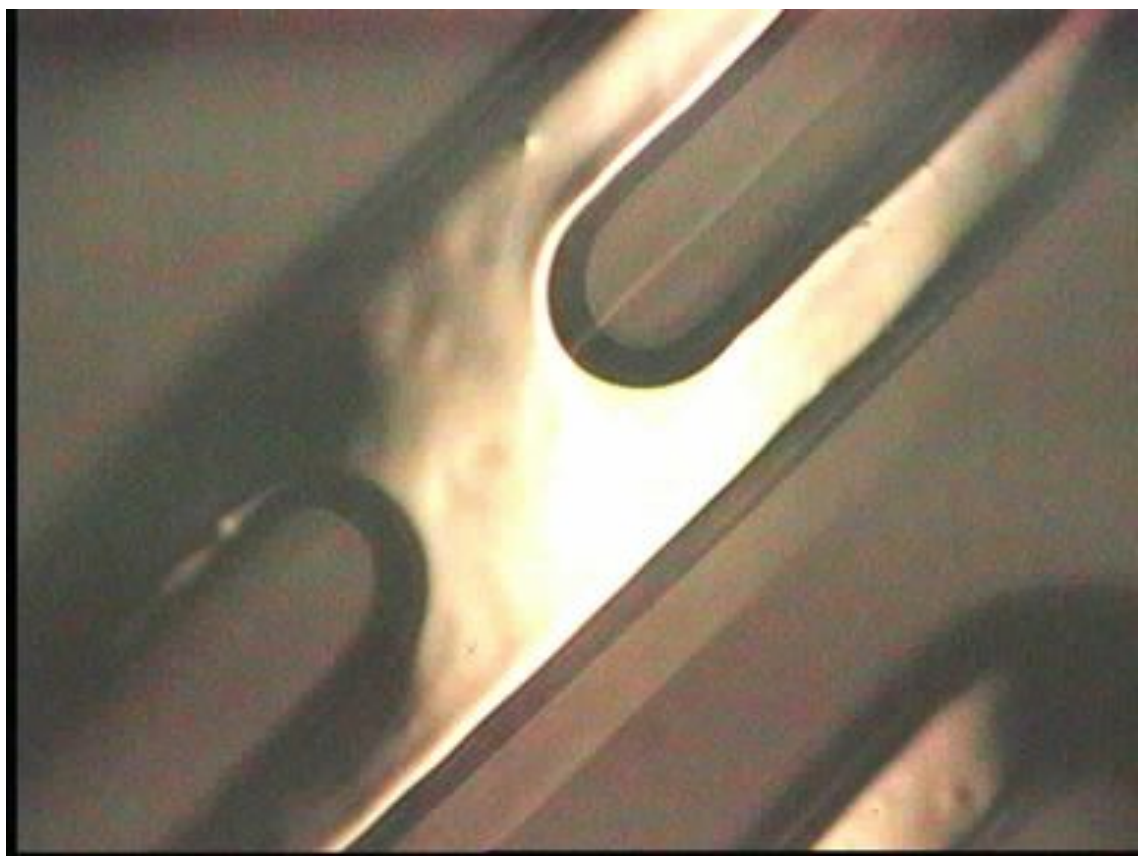
Initial stent. Size of photos is 1200x950 μm.
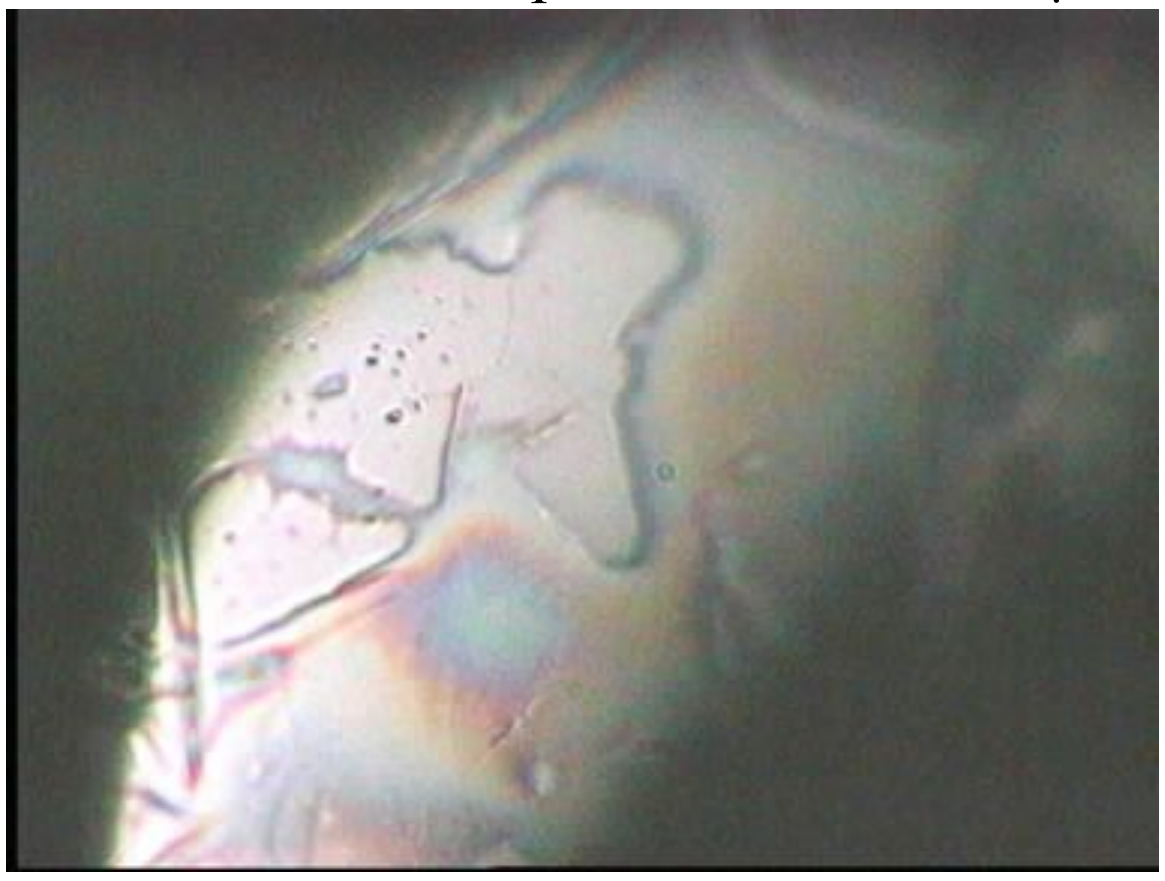
Untreated, after 10 days in 50ºC water. Size of photos is 240x190 μm.
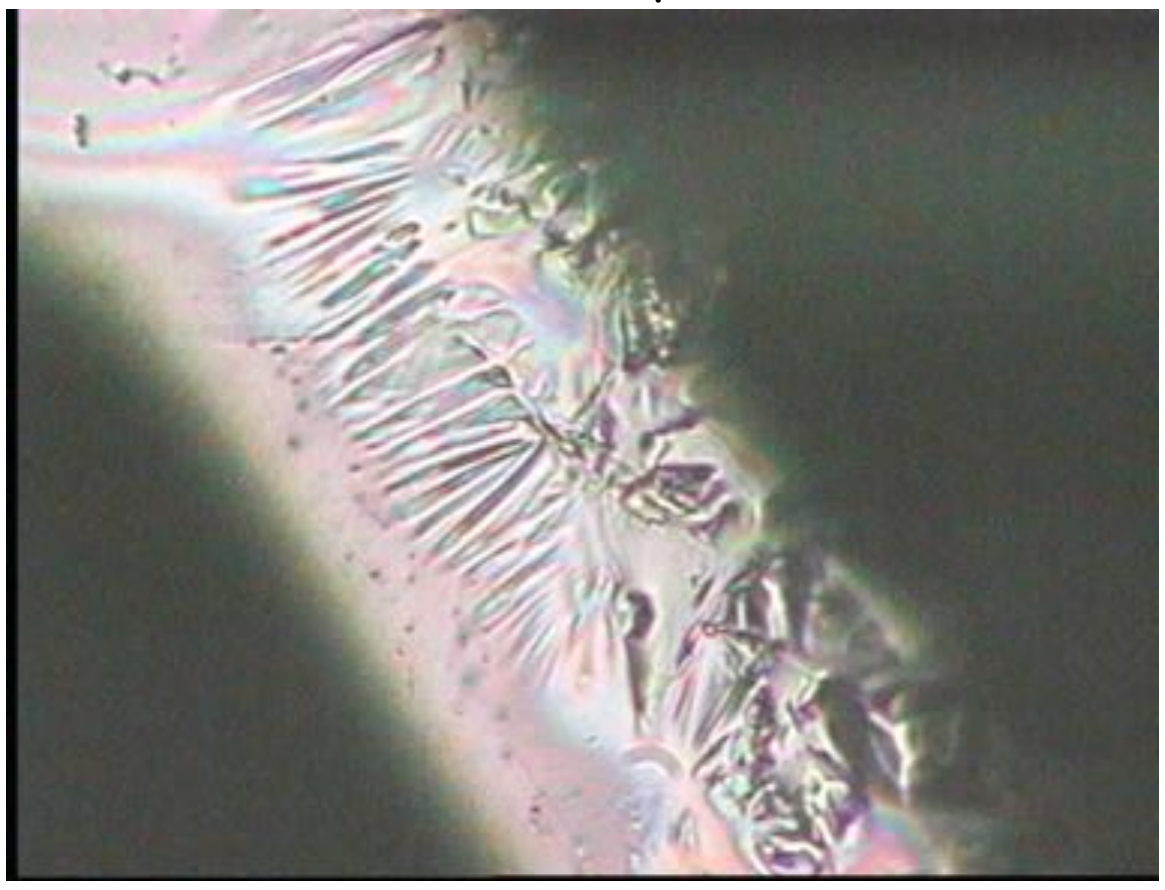
PIII treated, after 10 days in 50ºC water. Size of photos is 240x190 μm.

Fig.27. Microphotos of stents. The PLGA coating was treated by nitrogen ions of 20 keV energy and $5\times10^{15}$ ions/cm$^2$ fluence.